\documentclass[journal]{IEEEtran}
\usepackage{cite}

\ifCLASSINFOpdf
  \usepackage[pdftex]{graphicx}
  \DeclareGraphicsExtensions{.pdf,.jpeg,.png}
\else
  \usepackage[dvips]{graphicx}
  \DeclareGraphicsExtensions{.eps}
\fi
\usepackage[cmex10]{amsmath}
\usepackage{amssymb}

\usepackage{url}

\usepackage[usenames,dvipsnames]{color}
\definecolor{altncolor}{rgb}{0,0,0.8}
\usepackage[colorlinks=true, linkcolor=green, anchorcolor=altncolor,
citecolor=altncolor, filecolor=altncolor, menucolor=altncolor,
urlcolor=altncolor]{hyperref}

\usepackage{fancyheadings}

\begin{document}

\pagestyle{fancy}

\title{Validation of Advanced EM Models for UXO Discrimination}

%

\author{Peter B. Weichman}
%
\thanks{Peter Weichman is with BAE Systems, 6 New England Executive Park,
Burlington, MA 01803,  E-mail: peter.weichman@baesystems.com.}

\markboth{}%
{Weichman: Validation of advanced EM models for UXO discrimination}
%



\maketitle

%
%

\begin{abstract}

The work reported here details basic validation of our advanced
physics-based EMI forward and inverse models against data collected by
the NRL TEMTADS system. The data was collected under laboratory-type
conditions using both artificial spheroidal targets and real UXO. The
artificial target models are essentially exact, and enable detailed
comparison of theory and data in support of measurement platform
characterization and target identification. Real UXO targets cannot be
treated exactly, but it is demonstrated that quantitative comparisons
of the data with the spheroid models nevertheless aids in extracting
key target discrimination information, such as target geometry and
hollow target shell thickness.

\end{abstract}


%
\maketitle

\section{Introduction}
\label{sec:intro}

\IEEEPARstart{C}{leanup} of buried unexploded ordnance (UXO) from old
practice ranges is a longstanding economic and humanitarian problem.
Solution of the problem requires remote identification of subsurface
metallic objects. The most difficult technological issue is not the
detection of such targets---advanced metal detection systems, such as
the NRL-TEMTADS platform described below, easily detect even very small
amounts of metal at 1 m or more depths---but rather the ability to
distinguish between targets of interest and harmless clutter items,
such as various sized fragments of exploded ordnance. Since clutter
tends to exist at much higher density, even modest discrimination
ability leads to huge reductions in the economic cost of remediating
such sites \cite{SERDP}.

Formally, a successful solution to the electromagnetic (EM)
discrimination problem is an algorithm enabling accurate bounds on
physical properties of the target scatterer (position, shape,
orientation, composition, etc.)\ from measurements of the scattered
field using a well characterized platform (known transmitter/receiver
geometry, transmitted waveform, and so on). Solution of this inverse
problem requires a search over candidate solutions to the forward
problem, namely accurate forms for the scattered field from a known
target in a known subsurface environment. Generating high-fidelity
forward solutions requires full three dimensional numerical solutions
to the Maxwell equations, a difficult and time consuming computational
problem. To reduce the computational burden, it is extremely important
to obtain analytic solutions to as broad an array of exactly soluble
model problems as possible. These solutions may then either be used as
first-order models of the target, or as the basis of a perturbation
scheme for accurate modeling of ``nearby'' target geometries.

This paper details successful validation of our physics-based ``mean
field'' and ``early time'' approaches to modeling of time-domain
electromagnetic (TDEM) responses of compact, highly conducting targets.
Specifically, we apply our methods to the analysis (through both
forward and inverse modeling) of laboratory-style data collected by the
NRL TEMTADS system using artificial spheroidal targets, as well as some
real UXO targets. The models use the detailed measurement platform and
target parameters to generate highly numerically efficient, first
principles predictions for the measured time-domain voltages. The
models are designed to be essentially exact for spheroidal targets, and
the remarkable agreement between measurements and predictions strongly
supports this conclusion. The EM response of real UXO targets is found
to differ in significant ways from those of spheroids, but comparing
the two provides key insights into target identification.

The only compact targets for which a full analytic solution at any
frequency may be derived are those with spherical symmetry
\cite{Jackson}. These are rather poor approximations to UXO, which tend
to more resemble rounded cylinders or spheroids with roughly 4:1 aspect
ratio. Unlike for scalar wave problems, where exact solutions can be
generated also for ellipsoidal targets, the vector field Maxwell
equations fail to separate in ellipsoidal coordinates \cite{MF1953} and
fully analytic solutions do not exist.

As described in more detail below, the modeling approach applied here
uses simplifications available for UXO-like target shapes, and also in
different target electrodynamic regimes, to generate a combined
prediction that quantitatively describes the full response. We
specifically consider TDEM induction measurements. Here the transmitter
loop current pulse generates a magnetic field in the target region, and
this changing applied field, especially as the pulse terminates,
induces currents in the target, generating a scattered magnetic field.
The decaying scattered field, following pulse termination, induces the
measured voltage in the receiver loop.

In such a measurement there are three different regimes that one may
identify in the voltage time traces: early, intermediate, and late
time. At very early time, immediately following pulse termination, the
currents are confined to the immediate surface of the target. The
initial diffusion of these currents into the target interior leads to a
power law decay ($1/t^{1/2}$ for nonferrous targets, $1/t^{3/2}$ for
ferrous targets \cite{W2003,W2004}). At intermediate time, as the
currents penetrate the deeper target interior, the power law crosses
over to a multi-exponential decay, representing the simultaneous
presence of a finite set of exponentially decaying modes
\cite{WL03,W2011,W2012}. Finally, at late time only the single, slowest
decaying mode survives. We have developed a highly efficient
combination of analytic and numerical models, based on rigorous
solutions to the Maxwell equations, that covers these three regimes,
and the central purpose of this paper is to validate these models
against laboratory data from both artificial spheroidal and real UXO
targets, and to perform some inversion experiments that support their
use for target discrimination.

The outline of the paper is as follows. Details of the EM theory
underlying the models, and their numerical implementation, has been
presented elsewhere \cite{W2004,W2012}, but a basic overview is given
in Sec.\ \ref{sec:modelbg}. In Sec.\ \ref{sec:temtads} the basic
parameters of the NRL TEMTADS system are detailed. In Sec.\
\ref{sec:datacomp} model predictions are compared with TEMTADS data for
spherical targets, for which an exact analytic theory also exists
(Sec.\ \ref{sec:spheres}), and for prolate (elongated) and oblate
(discus-like) spheroidal targets (Sec.\ \ref{sec:spheroids}). In Sec.\
\ref{sec:inverse}, we describe results for the inverse problem, in
which various target properties are treated as unknown, and seek to
extract them from the data. In Sec.\ \ref{sec:realUXO} we describe
results for certain real UXO targets (specifically, 60 mm and 81 mm
mortar bodies). Finally, conclusions and directions for future work are
presented in Sec.\ \ref{sec:conclusions}.

\section{Modeling Background}
\label{sec:modelbg}

\subsection{Intermediate- to late-time modeling: mean field approach}
\label{sec:latertime}

Our approach to the intermediate and late time regimes is based on a
perturbation expansion about low frequency that takes advantage of the
fact that analytic solutions for ellipsoidal targets do exist in the
electrostatic limit (where the electric and magnetic fields are
gradients of scalar fields). Based on this, we have developed a
perturbation expansion about low frequency \cite{WL03,W2011,W2012} that
has an extremely efficient numerical implementation. The theory is
dubbed the ``mean field approach,'' since the expansion is highly
nonlocal in space, with the currents and fields at any given point in
the target being sensitive to their values throughout the target.
Although formally valid only at low frequency, the theory is extended
to higher frequencies by generating a large number of terms in the
series (for a related numerical approach using an expansion in
spheroidal wavefunctions, see also Refs.\ \cite{MIT1,MIT2,MIT3}).

For time-domain measurements, low frequency corresponds to later time,
in which initial rapid transients have died away. The solution to the
Maxwell equations allows one to represent the electric field following
pulse termination as a sum of exponentially decaying modes,
\begin{equation}
{\bf E}({\bf x},t) = \sum_{n=1}^\infty A_n
{\bf e}^{(n)}({\bf x}) e^{-\lambda_n t}
\label{2.1}
\end{equation}
where $\lambda_n$ are decay rates, ${\bf e}^{(n)}$ are mode shapes, and
$A_n$ are excitation coefficients. The first two are intrinsic
properties of the target, analogous to vibration modes of a drumhead.
Only the excitation amplitudes depend on the details of the measurement
protocol. At early time a very large number of exponentials is present,
and in fact the previously mentioned power laws arise from this large
superposition (see Sec.\ \ref{sec:earlyt} below).

As time progresses, modes with larger values of $\lambda_n$ decay more
quickly, and so at any given time $t$ the signal will be dominated by
some finite set of modes, namely those modes with $\lambda_n \lesssim
1/t$. At very late time, $t > 1/\lambda_1$, only the slowest decaying
mode contributes, and the signal becomes a pure exponential decay.
Thus, the earlier in time one wishes to model quantitatively, the
greater the number of modes that are required. The ultimate limitation
turns out to be the rate at which the excitation in pulse is
terminated. If the pulse is turned off on a time scale $t_r$ (see Sec.\
\ref{sec:tx_wvfrm}), then only modes with $\lambda_n \lesssim 1/t_r$
have substantial amplitudes $A_n$, and a finite set of modes suffices
for a full description of the target electrodynamics. For large
targets, this may require many thousands, or even tens of thousands of
modes, which is beyond current computational capability.

However, for computational purposes it is only required that enough
modes be computed that the resulting multi-exponential series overlaps
the early-time regime. The early-time power law and mode descriptions
may then be combined to fully describe the target dynamics over the
full measured time range. For modes that decay slowly enough, hence
contain low enough frequencies, the mean field approach can be used to
compute them, and compute as well the excitation level of each. We will
see that a few hundred modes is more than enough to attain the required
overlap, and this basically serves to define the beginning of what we
call the intermediate time regime.

Using the mode orthogonality relation (which follows from the Maxwell
equations),
\begin{equation}
\int d^3x \sigma({\bf x}) {\bf e}^{(m)*}({\bf x})
\cdot {\bf e}^{(n)}({\bf x}) = \delta_{mn},
\label{2.2}
\end{equation}
where $\sigma({\bf x})$ is the conductivity, the excitation amplitude
can be determined as
\begin{equation}
A_n = I_T^{(n)} N_T \int_{C_T} {\bf e}^{(n)*}({\bf x}) \cdot d{\bf l},
\label{2.3}
\end{equation}
in which the transmitter loop has been approximated by an ideal 1D loop
$C_T$ with $N_T$ windings, and
\begin{equation}
I_T^{(n)} = -\int_{-\infty}^0 dt e^{\lambda_n t} \partial_t I_T(t)
\label{2.4}
\end{equation}
depends on the history transmitter loop current $I_T(t)$ up until the
beginning of the measurement window, taken here as $t=0$. To gain some
intuition, a single perfect square wave pulse of amplitude $I_T^0$ and
duration $t_p$, one obtains
\begin{equation}
I_T^{(n)} = I_T^0 (1-e^{-\lambda_n t_p}).
\label{2.5}
\end{equation}
For a mode that decays rapidly on the scale $t_p$, one has $\lambda_n
t_p \gg 1$, and $I_T^{(n)} \simeq I_T^0$. For a more slowly decaying
modes, $I_T^{(n)}$ will have a strong dependence on $t_p$ and $n$. In
fact, for large targets one may actually encounter for small enough $n$
the regime $\lambda_n t_p < 1$ [e.g., $t_p = 25$ ms and $\tau_n =
1/\lambda_n = O(100\ \mathrm{ms})$] where $I_T^{(n)}$ will depend not
only on $t_p$, but on previous pulses.

Finally, the measured voltage takes the form
\begin{equation}
V(t) = \sum_{n=1}^\infty V_n e^{-\lambda_n t}
\label{2.6}
\end{equation}
in which, approximating the receiver as well by an ideal 1D loop $C_R$
with $N_R$ windings, the voltage amplitudes are given by the line
integrals
\begin{equation}
V_n = A_n N_R \int_{C_R} {\bf e}^{(n)}({\bf x}) \cdot d{\bf l}.
\label{2.7}
\end{equation}

Equations (\ref{2.3})--(\ref{2.7}) provide all the required ingredients
for generating predicted data based on a target and measurement
platform model. Our ``mean field'' numerical code divides naturally
into two parts.

The \emph{internal code} solves the Maxwell equations to produce the
intrinsic mode quantities $\lambda_n$ and ${\bf e}^{(n)}$ for a range
of expected targets. With increasing $\lambda_n$, the modes have more
complex spatial structure, and finite numerical precision means that
only a finite set (a few hundred) of slowest decaying modes are
actually produced \cite{W2011,W2012}.

The \emph{external} code uses the mode data, along with the measurement
platform data, to compute current integrals (\ref{2.4}), the line
integrals in (\ref{2.3}) and (\ref{2.7}), and then combines them to
output the voltage amplitudes $V_n$ and hence the time series
(\ref{2.6}). Note that the line integral computation requires full
knowledge of the relative position and orientation of the target and
platform.

For high precision, the internal code can take anywhere from minutes to
hours to produce mode data for a single target. However, given this
data, the external code takes at most a few seconds to produce the full
predictions. Precomputation and storage of a rapidly accessible
database of target data is therefore essential.

\begin{figure*}

\centerline{\includegraphics[width=2.33in]{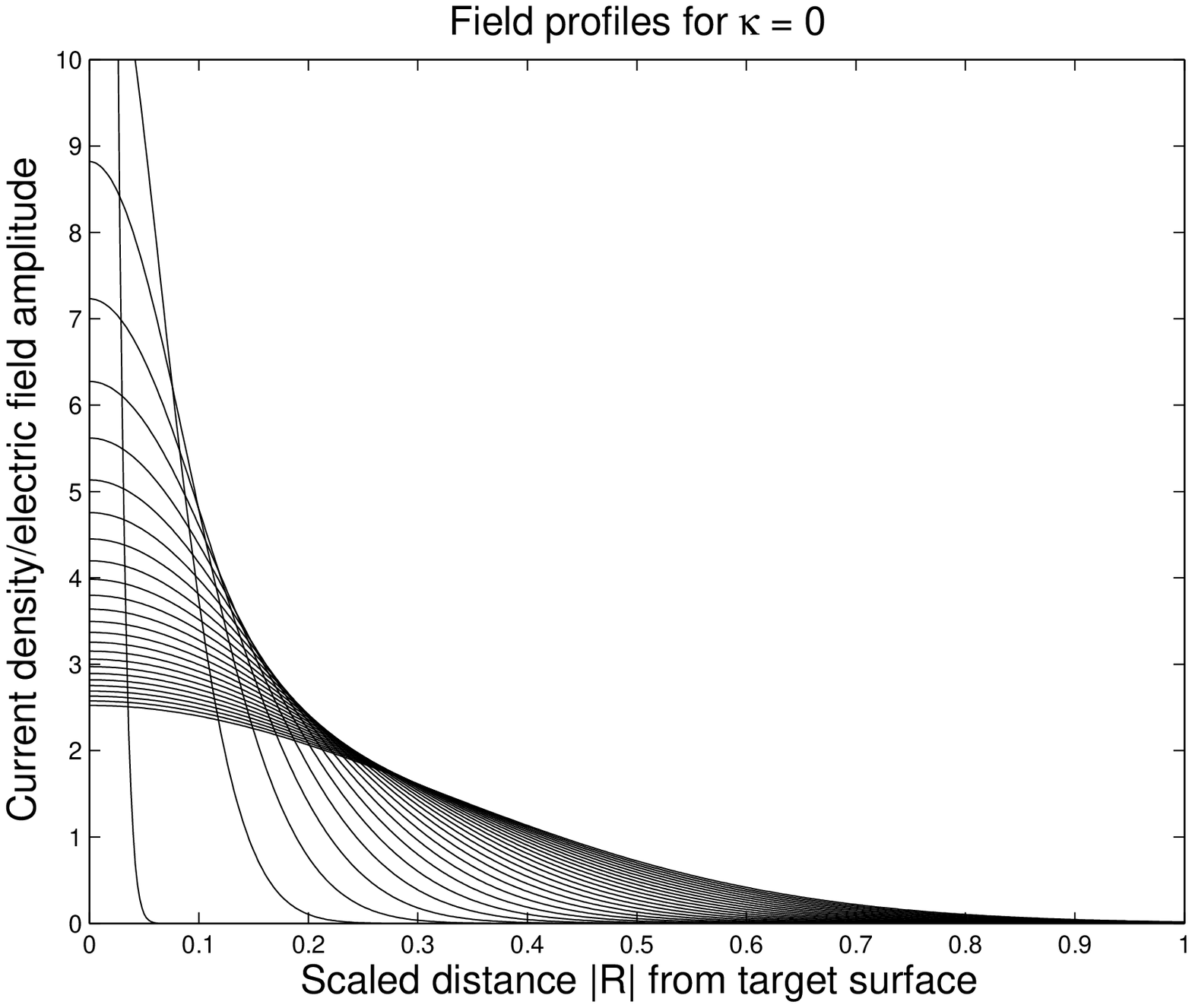}
\includegraphics[width=2.33in]{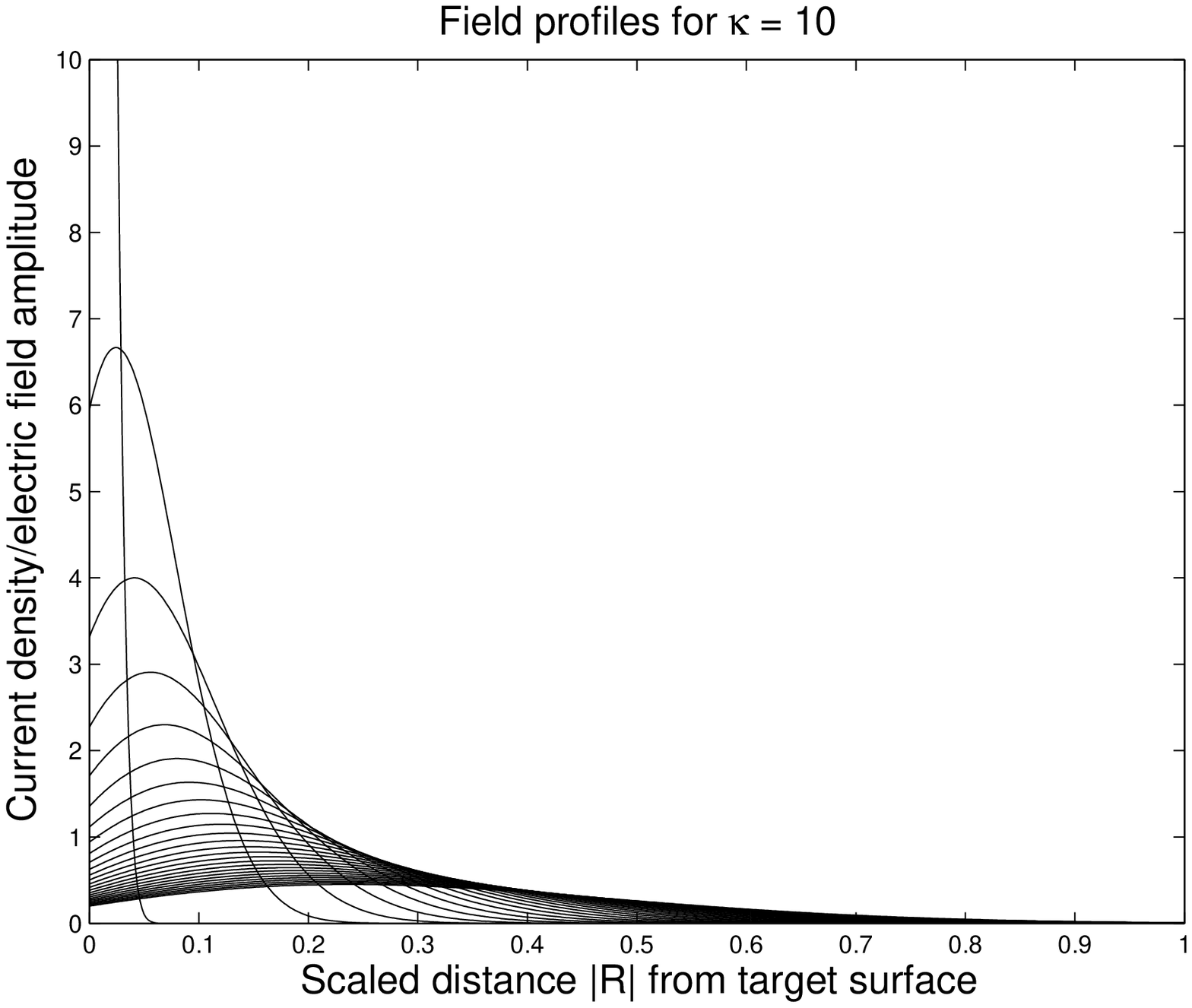}
\includegraphics[width=2.33in]{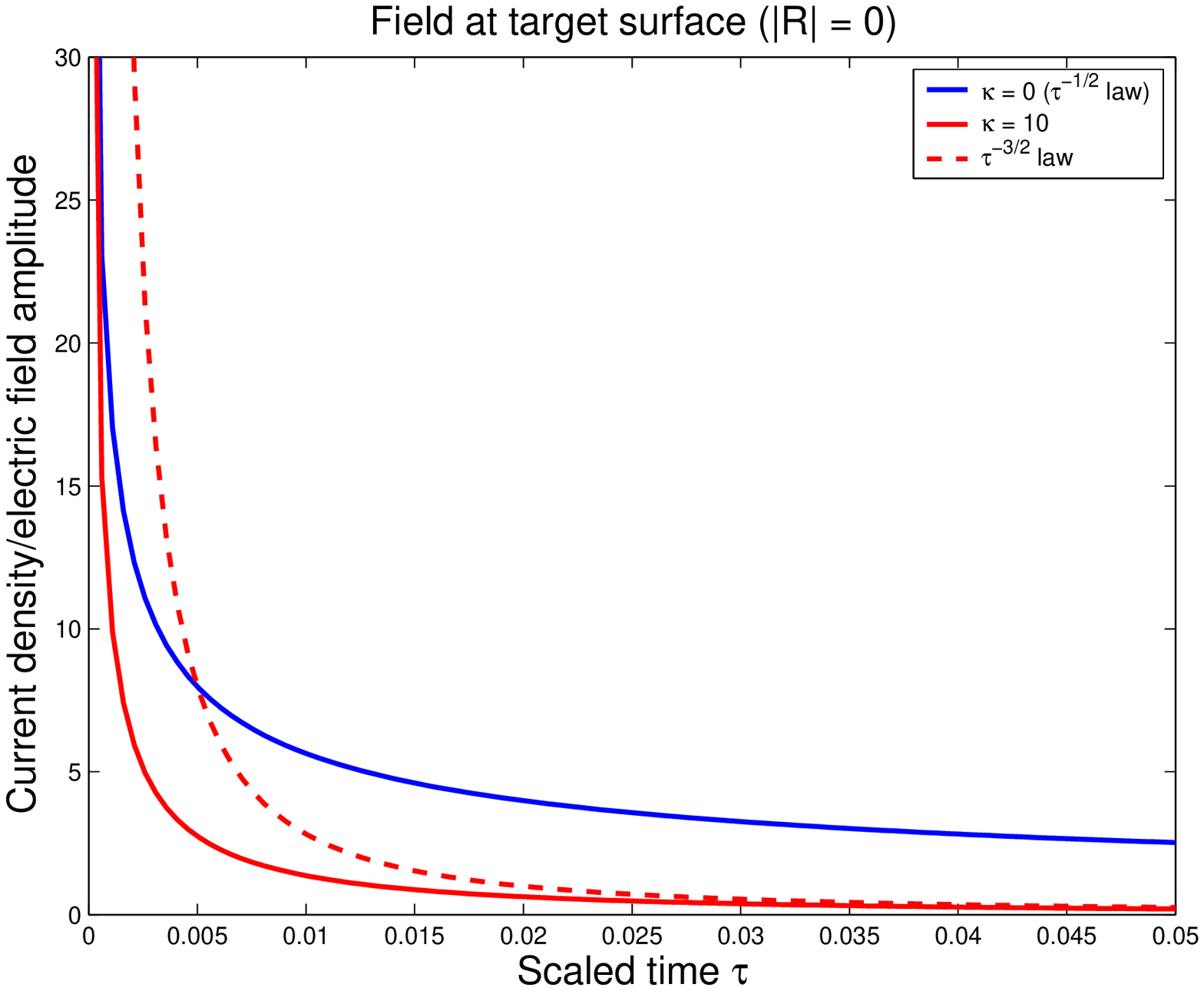}}

\caption{Illustration of the early time evolution of the surface
density depth profile from the target surface for nonmagnetic (left)
and magnetic (center) targets, beginning from a delta-function initial
condition (perfect step function pulse termination). Distance $R = r/L$
is scaled by the target size, time $\tau = t/\tau_D$ by the diffusion
time, so that $\kappa$ here corresponds $\kappa_n \sqrt{\tau_c}$ in
(\ref{2.11}), and is essentially the permeability contrast
$(\mu-\mu_b)/\mu_b$. The profiles are plotted for a sequence of 26
equally spaced scaled times $10^{-4} \leq \tau \leq 0.05$ (earlier
times corresponding to narrower profiles). The nonmagnetic profile
exhibits a pure Gaussian spreading into the target interior, while the
magnetic profile is much more complex due to the surface magnetic
boundary condition. Its maximum is pushed inwards from the boundary,
and decays more rapidly with time. The right plot shows the time trace
for the current density at the surface, $R=0$, and is essentially the
profile $H(\kappa \sqrt{\tau})$, equation (\ref{2.12}), which appears
in the measured voltage (\ref{2.11}). For $\kappa = 0$ (solid blue
line) the $\tau$-dependence follows an exact $1/\sqrt{\tau}$ power law.
For $\kappa > 0$ (solid red line) the $\tau$-dependence crosses over
from the identical $1/\sqrt{\tau}$  form at early-early time to the
$1/\tau^{3/2}$ power law (dashed red line) at late-early time [the
asymptotic forms displayed in (\ref{2.12})].}

\label{fig:earlyt_profiles}
\end{figure*}

\subsection{Complementary early time modeling}
\label{sec:earlyt}

For a rapidly terminated transmitter pulse, the external electric
field, and induced voltage, display an early time power law divergence
\cite{W2003,W2004} (saturating at very early time only on the scale of
the off-ramp time $t_r$. The boundary between the intermediate
(multi-exponential) and late time (mono-exponential) regime occurs at
the diffusion time scale
\begin{equation}
\tau_D = L^2/D
\label{2.8}
\end{equation}
where $L$ is the characteristic target radius, and $D = c^2/4\pi \mu
\sigma$ is the EM diffusion constant---this is the time scale required
for the initial surface currents to diffuse into the center of the
target. The early time regime corresponds to times $t \ll \tau_D$ (say,
$t < \tau_D/100$), beginning deep into the multi-exponential regime
where many (e.g., hundreds of) modes are excited. In this regime, for
nonpermeable, or weakly permeable targets ($\mu \simeq \mu_b$), one
obtains the simple power law prediction prediction \cite{W2003}
\begin{equation}
V(t) = V_e/t^{1/2},\ t \ll \tau_D,
\label{2.9}
\end{equation}
with all of the target and measurement parameters encompassed by the
single amplitude $V_e$, whose computation requires the solution of a
certain Neumann problem for the Laplace equation in the space external
to the target.

For permeable targets, a new magnetic time scale
\begin{equation}
\tau_\mathrm{mag} = \tau_D (\mu_b/\mu)^2
\label{2.10}
\end{equation}
emerges. For ferrous targets, $\mu/\mu_b = O(10^2)$, and
$\tau_\mathrm{mag}/\tau_c = O(10^{-4})$ is tiny. The early time voltage
then has a more complex \emph{magnetic surface mode} structure,
\begin{equation}
V(t) = \sum_{n=1}^\infty V_n^e H(\kappa_n \sqrt{t})
\label{2.11}
\end{equation}
where the $\kappa_n$ are surface mode eigenvalues, and the mode time
trace profile
\begin{eqnarray}
H(s) &=& \frac{1}{\sqrt{\pi} s} - e^{s^2} \mathrm{erfc}(s)
\nonumber \\
&\approx& \left\{\begin{array}{ll}
\frac{1}{\sqrt{\pi} s}, & s \ll 1 \\
\frac{1}{2\sqrt{\pi}s^3}, & s \gg 1,
\end{array} \right.
\label{2.12}
\end{eqnarray}
where $\mathrm{erfc}(s)$ is the complementary error function,
interpolates between a $1/t^{1/2}$ power law at early-early time, $t
\ll \tau_\mathrm{mag}$, and a $1/t^{3/2}$ power law at late-early time,
$\tau_\mathrm{mag} \ll t \ll \tau_D$. For large ferrous targets, this
latter interval is very large, and may, in fact, accurately represent
the signal over nearly the entire measurement interval (see Sec.\
\ref{sec:datacomp}).

Figure \ref{fig:earlyt_profiles} illustrates the important features of
the early time modeling, including the complex evolution of the surface
current depth profile [which extends $H(s)$ to a function of both time
and space \cite{W2004}] that ultimately gives rise to the externally
measured voltage (\ref{2.11}).

The surface modes are special surface current profiles (two such
patterns are shown in Fig.\ \ref{fig:earlytmodes} below) that, instead
of decaying exponentially, evolve according to the universal function
$H(s)$. They and the $\kappa_n$ are solutions to an eigenvalue problem
defined on the surface of the target \cite{W2004}. They may be
determined analytically only for spherical targets, where one finds
\begin{equation}
\kappa_l = l/\sqrt{\tau_\mathrm{mag}},\ l=1,2,3,\ldots,
\label{2.13}
\end{equation}
each $(2l+1)$-degenerate, with $\tau_\mathrm{mag} = 4\pi \sigma \mu_b^2
a^2/\mu c^2$, where $a$ is the radius. The surface current patterns are
controlled by the spherical harmonics of order $l$. The amplitudes
$V_n^e$ again require a solution to an external Laplace-Neumann
problem.

Unlike the bulk, exponential modes, under most conditions, only a very
few surface modes are excited. The initial surface current pattern
more-or-less follows the shape of the magnetic field generated by the
transmitter coil. Unless the target is close to the coil, this field is
fairly uniform, and the corresponding surface current density is fairly
uniform as well, and can then be represented by the first few (two or
three) modes. There is a very heavy numerical overhead in computing
these modes and their excitation amplitudes, all in pursuit of
predicting the rather limited information content of just a few
coefficients. Given the success of extending the mean field predictions
into the intermediate-early time regime, we have therefore found that
it is much more efficient to extend the voltage curve by \emph{fitting}
the data at intermediate times to a one or two term series of the form
(\ref{2.11}), estimating $\kappa_n \approx 1/\sqrt{\tau_\mathrm{mag}}$
for the first few modes. Although this precludes quantitative
predictions at early-early time, it provides an enormously useful
\emph{qualitative} confirmation that the functional form $H(s)$
accurately describes the data.

\begin{figure}

\centerline{\includegraphics[width=2.6in]{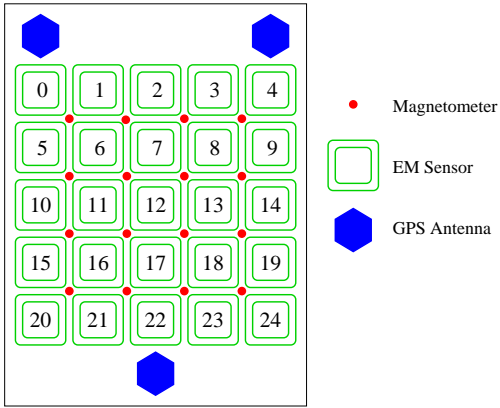}}

\caption{Sketch of NRL TEMTADS array consisting of a $5 \times 5$ array
of 25 independent, concentric transmitter and receiver coils, numbered
from 0 to 24 as shown. Due to rapid decay of signals with target depth,
precise (cm level) geometry and placement of the coils (summarized in
Table \ref{table:temtads_array}) can have significant effect on the
overall measured voltage amplitude.}

\label{fig:temtads_array}
\end{figure}

\begin{table}

\begin{center}
\begin{tabular}{lll}
Sensor center horizontal separation       & & 40 cm \\
Transmitter coil center height            & & 4.3 cm \\
Transmitter diameter                      & & 35 cm \\
Number of transmitter coil windings $N_T$ & & 35 \\
Receiver coil center height               & & 0.4 cm \\
Receiver diameter                         & & 25 cm \\
Number of receiver coil windings $N_R$    & & 16 \\
\end{tabular}
\end{center}

\caption{NRL TEMTADS array geometry. The transmitter coil windings are
7.8 cm tall with 0.4 cm thick endcaps on top and bottom. Height is
measured from the bottom side of the lower endcap, and the transmitters
are then modeled as an idealized 1D square loops at $0.4 + 3.9 = 4.3$
cm height. The receiver coils are vertically compact and lie at the
bases of the transmitter coils, hence are modeled as idealized 1D
square loops at 0.4 cm height.}

\label{table:temtads_array}
\end{table}

\section{TEMTADS platform}
\label{sec:temtads}

\subsection{Platform geometry}
\label{sec:pltfrm_geom}

The $5 \times 5$ NRL TEMTADS sensor array is sketched in Fig.\
\ref{fig:temtads_array}, and its geometrical parameters are summarized
in Table \ref{table:temtads_array}. The loops $C_T$ and $C_R$ are all
modeled as perfect squares with 35 cm and 25 cm edges, respectively.
The origin is taken to be at the base of the lower endcap for sensor
12, the positive $x$-axis towards sensor 13, the positive $y$-axis
towards sensor 7, and the positive $z$-axis vertically upwards. The
transmitter and receiver loop centers then all have $x$- and
$y$-coordinates that are multiples of 40 cm. The transmitters are all
at $z=4.3$ cm, and receivers are all at $z=0.4$ cm. Target positions
and orientations quoted in later sections are all defined relative to
this frame of reference.

The precise overall voltage amplitudes, required at least for initial
verification of the instrument calibration, turn out to be surprisingly
sensitive to small changes in these numbers. The scattered fields are
approximately dipolar, and the voltage therefore decreases roughly as
$1/d^6$ with depth $d$. For example, therefore, a 1 cm error for a 30
cm deep target then leads to a 20\% error in the voltage amplitude. A
consistent systematic error of this magnitude, in fact, is what led us
to discovering the existence of the endcaps, and the vertical offset
between the transmitter and receiver loops.

\begin{figure}

\centerline{\includegraphics[width=3.2in,bb=70 160 535 640,clip]
{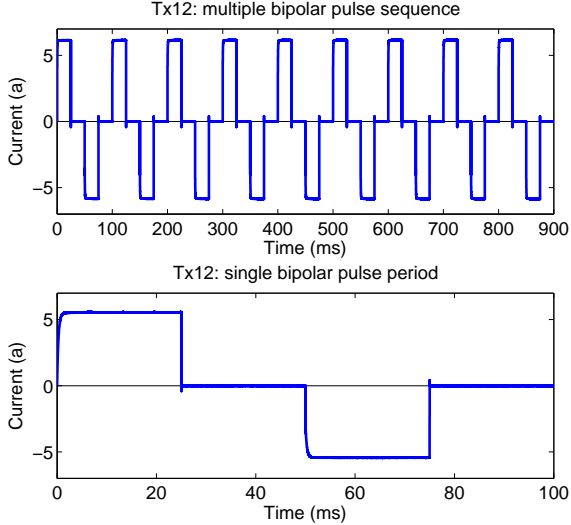}}

\caption{TEMTADS transmitter current bipolar pulse waveform.
\textbf{Top:} multiple periods. \textbf{Bottom:} single 100 ms period.}

\label{fig:pulse_global}
\end{figure}

\subsection{Transmitter waveform}
\label{sec:tx_wvfrm}

The TEMTADS bipolar pulse sequence is shown in Fig.\
\ref{fig:pulse_global}. Each pulse is $t_p = 25$ ms long, followed by a
25 ms measurement window.  An adequate model of the pulse waveform is
the form:
\begin{equation}
I(t) = \left\{\begin{array}{ll}
I_\mathrm{max} (1 - e^{-t/\tau_1}), & 0 < t \leq t_p
\\
I_\mathrm{max} [1-(t-t_2)/t_r], & t_2 < t \leq t_p+t_r
\end{array} \right.
\label{3.1}
\end{equation}
with exponential onset time constant $\tau_1 = 0.33$ ms, off-ramp time
$t_r = 10\ \mu$s, and current amplitude $I_\mathrm{max} = 5.7 \pm 0.3$
a. This form misses some detailed multi-exponential behavior during the
pulse onset that can be shown to have negligible effect on the
excitation coefficients. The second half of the full bipolar pulse,
beginning at $t = 2t_p$, is the same as the one above, but inverted.
The functional forms in (\ref{3.1}) are simple enough that analytic
forms for the current coefficients (\ref{2.4}) may be computed
straightforwardly.

\begin{figure}

\centerline{\includegraphics[width=3.0in,bb=108 207 477 585,clip]
{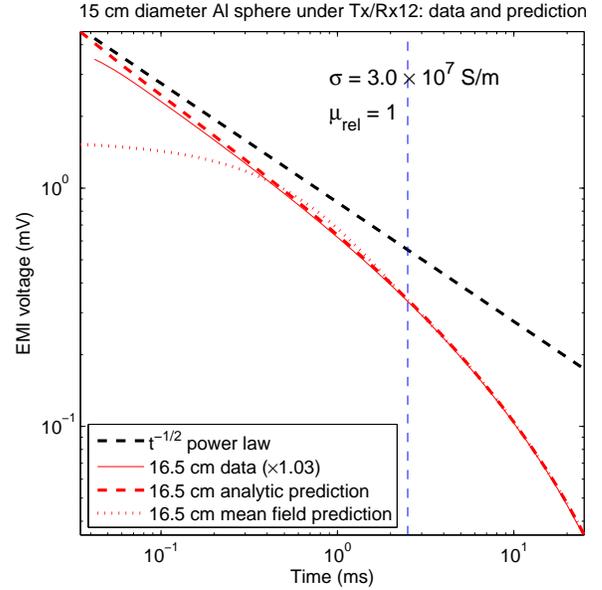}}

\caption{Data and theory for a 15 cm diameter aluminum sphere with
center lying 16.5 cm below the center of sensor 12 (see Fig.\
\ref{fig:temtads_array}), which is also the only active sensor. The
solid red line is the data, the dashed red line the prediction from the
exact analytic solution for the sphere, the dotted red line is the mean
field prediction (based on 232 modes), and the dashed black line is the
early time $1/\sqrt{t}$ power law. The 1.03 overall multiplier listed
in the legend has been applied to the data to optimize the fit, and is
well within the expected 10\% fluctuation in the TEMTADS current
amplitude. The vertical dashed line marks the rough division between
the early time and multi-exponential ($\lesssim 100$ modes) regimes,
and it is seen that the mean field prediction is valid well into the
early time regime. The slight deviation of the data from the analytic
prediction at very early time, $t < 0.1$ ms, is likely an instrument
saturation effect (seen much more clearly in Fig.\ \ref{fig:stsphcomp},
beginning roughly at the same voltage level).}

\label{fig:alsphcomp}
\end{figure}

\begin{figure}

\centerline{\includegraphics[width=3.0in,bb=71 177 515 616,clip]
{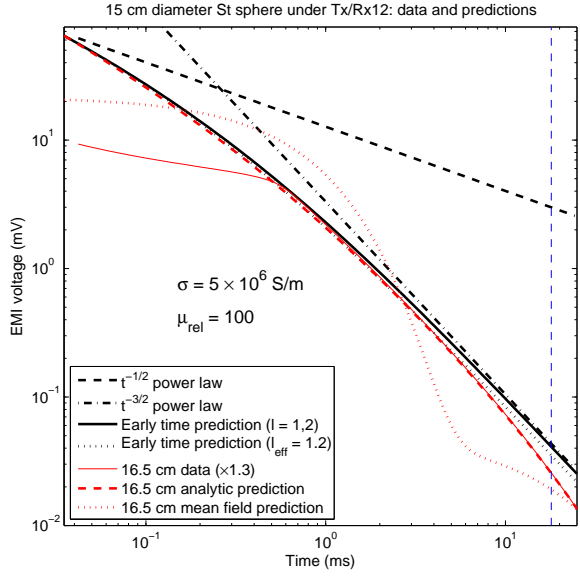}}

\caption{Data and theory for a 15 cm diameter steel sphere with center
lying 16.5 cm below the center of sensor 12 (see Fig.\
\ref{fig:temtads_array}), which is also the only active sensor. The
solid red line is the data (which shows clear instrument saturation
above a few volts), the dashed red line the prediction from the exact
analytic theory solution for the sphere, the dotted red line is the
mean field prediction (based on 232 modes). The dashed black line is a
two term fit to the early time form (\ref{2.11}) using the known values
(\ref{2.13}), and the dotted black line is a single term fit using
$\kappa_1$ as a fit parameter. The vertical dashed line marks the rough
division between the early time and multi-exponential ($\lesssim 100$
modes) regimes, and is much later here than in Fig.\
\ref{fig:alsphcomp} because the EM time scale is proportional to the
product $\sigma \mu$, which is an order of magnitude larger here. For
reasons described in the text, the mean field prediction has a more
complex structure for ferrous targets, and penetrates only to the edge
of the early time regime (it is the fact that it is accurate beyond
about 20 ms that is the real figure of merit here, as would be more
evident if the data extended to later time). The 1.3 multiplier listed
in the legend is that applied to the data to optimize the fit, and lies
outside the expected 10\% fluctuation in the current amplitude. The
difference is likely the result of small positioning errors. Sensor
saturation is apparent below about 0.5 ms. The late-early time
$1/t^{3/2}$ power law is evident in the data, but full convergence to
the $1/\sqrt{t}$ early-early time power law is incomplete, and not
expected until about 10 $\mu$s.}

\label{fig:stsphcomp}
\end{figure}

\section{Data comparisons}
\label{sec:datacomp}

\subsection{Spherical targets}
\label{sec:spheres}

Having described the electromagnetic model, and the platform model
required to implement it, we now turn to its validation with real data.
We begin with spherical targets, for which exact analytic solutions
exist in both the early time \cite{W2003,W2004} and multi-exponential
regimes \cite{Jackson}. This allows one to validate the sensor model
under conditions where the target model is fully specified.

Figure \ref{fig:alsphcomp} shows results for a 15 cm diameter aluminum
sphere, plotted on both linear and log time scales---the latter much
more clearly verifies the asymptotic $1/\sqrt{t}$ early time power law.
The agreement is quite remarkable---note that the vertical scale is in
millivolts, not an arbitrary scaled unit. The only real fitting
parameter is the conductivity, and the chosen value $\sigma = 3 \times
10^7$ S/m is well within the range expected for aluminum. As discussed
in Sec.\ \ref{sec:temtads}, the overall pulse-to-pulse transmitter
current amplitude is stable only at the 10\% level. This leads to an
identical uncertainty in the overall voltage amplitude. In the figure,
an overall factor of 1.03 has been applied to the data to obtain an
optimal fit, well within this uncertainty. The slowest decaying mode
for this target is $\tau_1 = 21.5$ ms, so the measurement window here
barely enters the late time regime $t \gtrsim \tau_1$. The mean field
prediction, based on an approximate calculation of the first 232 modes
\cite{W2011,W2012}, is seen to accurately describe the data well into
the early time regime.

The mean field prediction has much more interesting structure for
ferrous targets. Due to the nature of the EM boundary conditions in the
large permeability contrast limit, rather than computing only the
slowest decaying modes, two distinct sets of slow (169 modes in this
case, with time constants larger than 3.01 ms) and fast (63 modes in
this case, with time constants smaller than 0.74 ms) decaying modes are
produced, with large gap between that would only be filled if one
pushed the computation to higher order. This is the source of the
S-curve-like structure seen in the right panel of Fig.\
\ref{fig:stsphcomp}. The reduction in the number of slowly decaying
modes reduces the accuracy of the theory near the early--intermediate
time boundary (as compared to the nonmagnetic case shown in Fig.\
\ref{fig:alsphcomp}), but the presence of the more rapidly decaying
modes at least provides an improved trend at very early time. The
slowest decaying mode for this target has a time constant $\tau_1 =
180$ ms, indicating a late time regime an order of magnitude beyond the
measurement.

The early time prediction, which follows both the exact solution and
the data over a significant fraction of the time interval, deserves
some comment. As described in Sec.\ \ref{sec:earlyt}, to obtain the
solid black lines in Fig.\ \ref{fig:alsphcomp}) the known eigenvalues
(\ref{2.13}) are used, but the amplitudes $V_n^e$ are determined
(\ref{2.11}) by fitting to the data. Only two terms are kept,
\begin{equation}
V(t) = V_0 \left[(1-\alpha) H\left(\sqrt{t/t_\mathrm{mag}} \right)
+ \alpha H\left(2\sqrt{t/t_\mathrm{mag}} \right) \right]
\label{4.1}
\end{equation}
with the known value $t_\mathrm{mag} = 0.35$ ms, and the amplitude $V_0
= 83$ V, and mixing parameter $\alpha = 0.4$ are fit. The one term
series $V_0 = 60$ V, $\alpha = 0$ provides an adequate, but lower
quality fit.

However, a better fit than both of these is provided by a single term
series in which one allows the eigenvalue $\kappa_1$ to be adjusted.
The dotted black line in Fig.\ \ref{fig:alsphcomp}) shows the result
obtained using $\kappa_1 = l_\mathrm{eff}/\sqrt{t_\mathrm{mag}}$ with
$l_\mathrm{eff} = 1.2$, along with amplitude $V_1^e = 80$ V. This will
be our fitting method of choice for non-spherical targets, where the
eigenvalues $\kappa_n$ have not yet been computed.

It is worth emphasizing the importance of the fact that analytic
functional forms of the type (\ref{4.1}) fit the data so well. The
log-time plot demonstrates that the data span the full range over which
the argument $s$ in (\ref{2.12}) interpolates between the two power
laws \cite{foot:multiscale}. The data therefore has significant
structure through this time range, but this does not reflect any deep
structure of the target (beyond the fact that it is ferrous). Quite the
contrary: as illustrated in Fig.\ \ref{fig:earlyt_profiles} it
represents the dynamics of a laterally very smooth surface current
sheet as it begins to penetrate the first centimeter or so into target.
The complexity arises strictly from the interplay between the electric
and magnetic field boundary conditions at the surface. This serves to
confirm that the early time regime provides limited target
discrimination ability (again, beyond the fact that it is ferrous).

\begin{figure}

\centerline{\includegraphics[width=3.2in,bb=20 125 590 675,clip]
{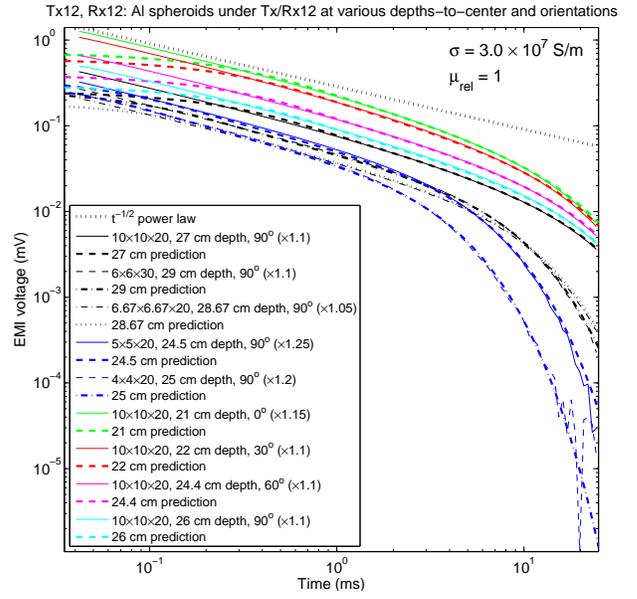}}

\caption{Consolidated plot of data and theory for a range of artificial
aluminum prolate spheroidal targets, centered under Tx/Rx12 at various
depths and orientations. The dimensions listed in the legend are
diameters. Orientation angles indicate symmetry axis declination
(toward the center of sensor 11), so that $0^\circ$ corresponds to
vertical and $90^\circ$ to horizontal. The multipliers are again the
overall factors applied to the data to obtain optimal agreement with
the prediction. The thick dashed lines are the mean field predictions,
which show remarkable agreement well into the early time regime, where
the onset of the $1/\sqrt{t}$ power law is evident.}

\label{fig:alprolate}
\end{figure}

\begin{figure}

\centerline{\includegraphics[width=3.2in,bb=-25 85 620 715,clip]
{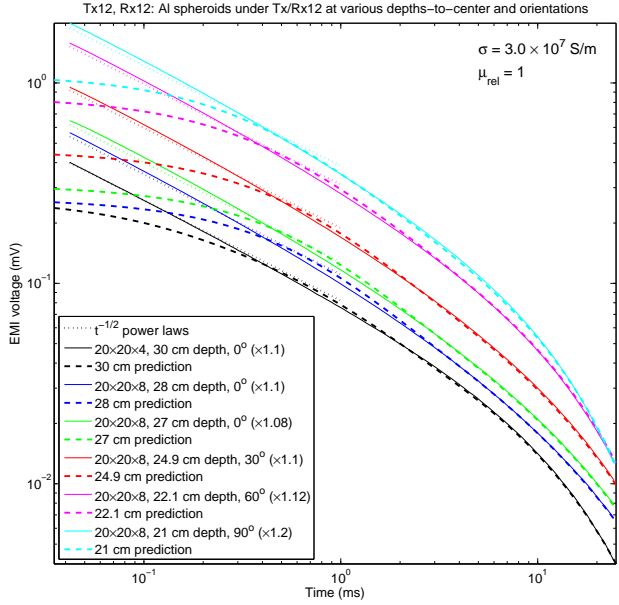}}

\caption{Consolidated plot of data and theory for a range of artificial
aluminum oblate spheroidal targets, centered under Tx/Rx12 at various
depths and orientations. The thick dashed lines are the mean field
predictions, which show remarkable agreement well into the early time
regime, where the $1/\sqrt{t}$ power law takes over (dotted lines).}

\label{fig:aloblate}
\end{figure}

\subsection{Prolate and oblate spheroidal targets}
\label{sec:spheroids}

Having verified instrument calibration and several other quantitative
details under conditions where an exact solution exists, results for
spheroidal targets are now presented.

Figure \ref{fig:alprolate} shows a consolidated plot of data and theory
for various prolate (elongated) spheroidal aluminum targets at various
depths and orientations. Spheroid aspect ratios $a_z/a_{xy}$ vary
between 2 and 5.

The theoretical plots (thick dashed lines) are the mean field
predictions based on the first 232 modes. It is again evident that the
mean field predictions are valid well into the early time regime. The
multi-exponential time series eventually saturates and falls below the
data, but not before the $1/\sqrt{t}$ power law is quite well
established. For smaller targets (e.g., the $4 \times 4 \times 20$ cm
and $5 \times 5 \times 20$ cm spheroids) the mean field prediction can
cover nearly the entire measurement window. The multi-exponential time
series eventually saturates and falls below the data, but not before
the $1/\sqrt{t}$ power law is quite well established. Interpolating
between the mean field prediction and this power law clearly enables
one to accurately match the data over the full range (this will be
demonstrated more quantitatively in the inversion experiments described
below).

It is apparent that most of the target discrimination information
occurs at intermediate to late time. The traces are all more-or-less
parallel at early time, and variations in the overall amplitude from
from variation in depth, size or geometry of the target are not
distinguishable. On the other hand, at later time, the traces for
smaller targets (e.g., again, the $4 \times 4 \times 20$ cm and $5
\times 5 \times 20$ cm spheroids) drop off much more quickly than those
of larger targets.

There are also interesting dependencies on target orientation in this
regime (green, red, magenta, and cyan curves in the upper part of the
plot for the $10 \times 10 \times 20$ cm spheroid \cite{foot:orient}).
For a vertical target, the excited modes are dominated by currents that
circulate around the symmetry axis, while for a horizontal target the
currents tend to circulate along it. The horizontal target mode has a
slower decay rate (time constant $\tau_h = 13.7$ ms vs.\ $\tau_v =
12.0$ ms), and couples differently to the transmitted field, and this
is visible in the later-time traces.

Identical conclusions are evident from the data on oblate (discus-like)
spheroidal aluminum targets (aspect ratios $\alpha = 0.2, 0.4$) shown
in Fig.\ \ref{fig:aloblate}. Here we have overlayed segments of
$1/\sqrt{t}$ power law on each curve, explicitly demonstrating
successful interpolation (with, perhaps, 5--10\% errors in the overlap
regime).

\begin{figure}

\centerline{\includegraphics[width=3.2in,bb=-80 55 670 750,clip]
{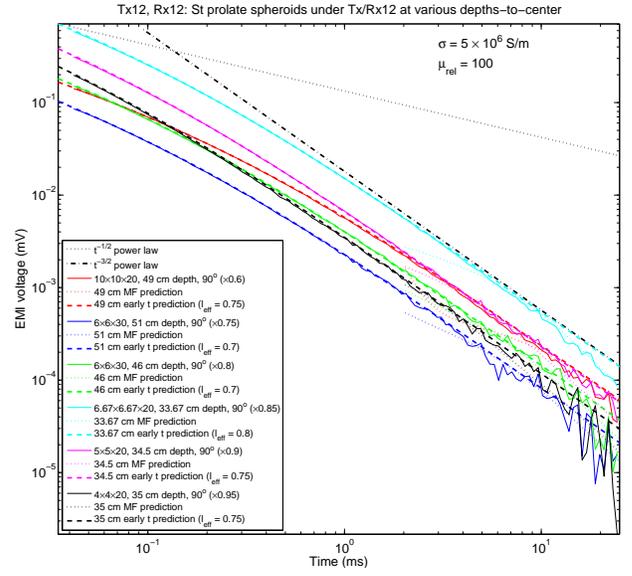}}

\caption{Consolidated plot of data and theory for a range of steel
prolate spheroidal targets, centered under Tx/Rx12 at various depths
and orientations. The thick dashed lines are the early time
predictions, which show remarkable agreement over nearly the entire
measurement window. The latter take the form (\ref{2.11}) with a
\emph{single} term, in which the amplitude $V_1^e$ and eigenvalue
$\kappa_1 = l_\mathrm{eff}/\sqrt{\tau_\mathrm{mag}}$ are adjusted to
optimize the fit. Here $\tau_\mathrm{mag}$ is defined by (\ref{2.10})
and (\ref{2.8}), with the choice $L = \min\{a_{xy},a_z\} = a_{xy}$. The
mean field predictions are shown by the dotted lines. If extended over
the full time interval, they also would display the S-curve behavior
seen in Fig.\ \ref{fig:stsphcomp}. For these smaller targets, their
region of validity begins only at later times where the signal levels
are falling into the noise floor. The early time regime therefore
encompasses almost the full range of useful data.}

\label{fig:stprolate}
\end{figure}

\begin{figure}

\centerline{\includegraphics[width=3.2in,bb=40 165 555 630,clip]
{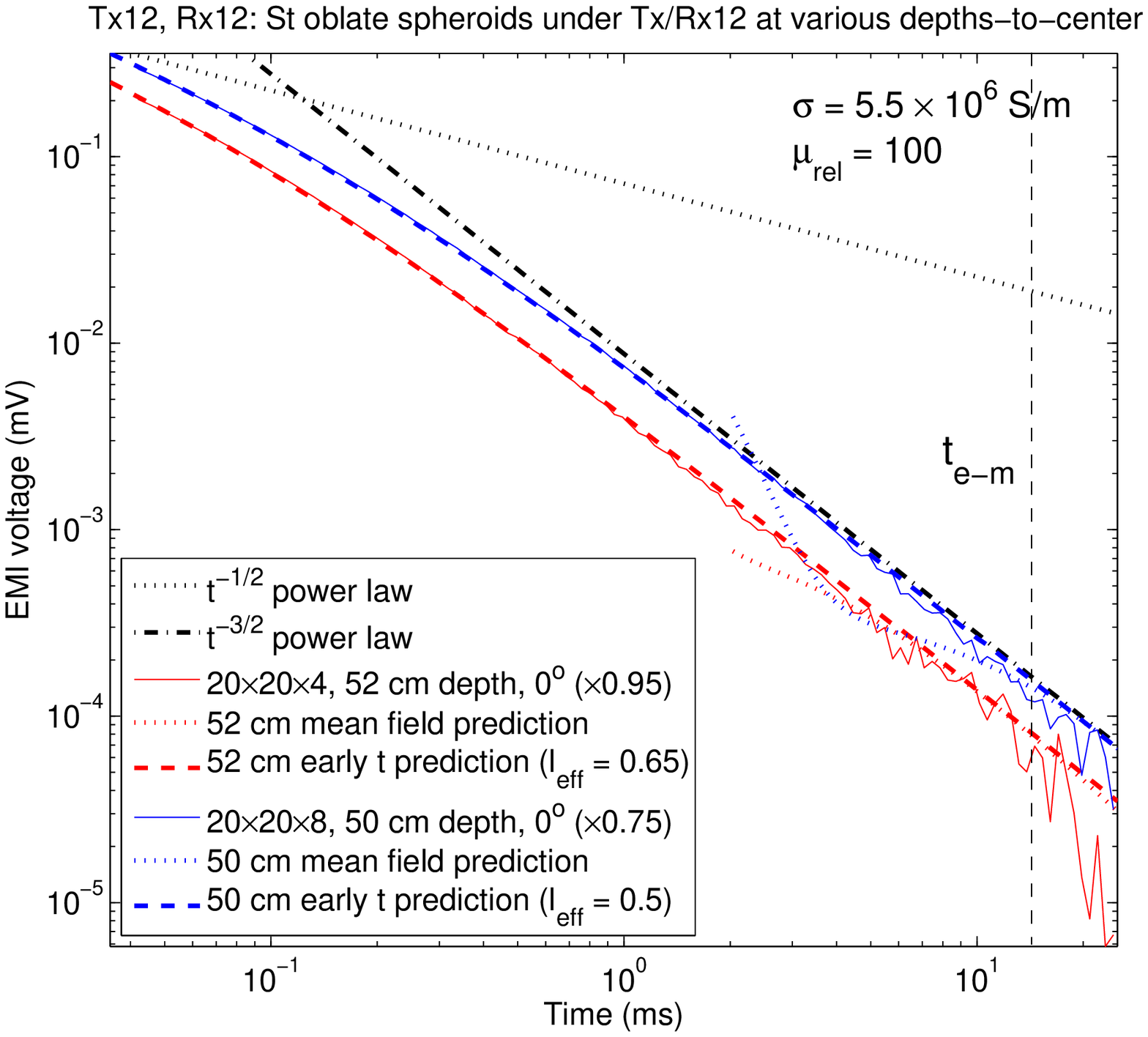}}

\caption{Consolidated plot of data and theory for a range of steel
oblate spheroidal targets, centered under Tx/Rx12 at various depths and
orientations. The thick dashed lines are the early time predictions,
which show remarkable agreement over nearly the entire measurement
window. The latter take the form (\ref{2.11}) with a \emph{single}
term, in which the amplitude $V_1^e$ and eigenvalue $\kappa_1 =
l_\emph{eff}/\sqrt{\tau_\mathrm{mag}}$ are adjusted to optimize the
fit. Here $\tau_\mathrm{mag}$ is defined by (\ref{2.10}) and
(\ref{2.8}), with the choice $L = \min\{a_{xy},a_z\} = a_z$. The mean
field predictions are shown by the dotted lines. For these smaller
targets, their region of validity again begins only as the signal
levels are falling into the noise floor. The early time regime ($t <
t_\mathrm{e\mbox{-}m}$) encompasses nearly the full range of useful
data.}

\label{fig:stoblate}
\end{figure}

The dependence on orientation is much stronger for oblate spheroids
(green, red, magenta, and cyan curves for the $20 \times 20 \times 8$
cm spheroid \cite{foot:orient}). Because it is being ``squeezed''
vertically, the horizontal target (discus on edge) mode now has
significantly faster decay rate than vertical target (discus lying
flat) mode (time constant $\tau_h = 13$ ms vs.\ $\tau_v = 24$ ms).
Because the the latter mode is not excited at all when the target is
horizontal, the $90^\circ$ (cyan) curve in Fig.\ \ref{fig:aloblate}
decays much more quickly at late time than the other curves.

In both Figs.\ \ref{fig:alprolate} and \ref{fig:aloblate} the
multipliers used to scale the data for optimal fit appear to have a
small ($\sim 10$\%) systematic bias that cannot be explained by random
variation in the transmitter loop current. A combination of small
conductivity and positioning errors is the likely culprit.

Figures \ref{fig:stprolate} and \ref{fig:stoblate} show data and theory
for steel prolate and oblate spheroidal targets. As for the steel
sphere (Fig.\ \ref{fig:stsphcomp}), the early time regime dominates,
and the mean field results (dotted curves; with S-curve behavior
excised in this case so as not to confuse the plots) are valid only
over a small part of the time interval where the data is already
becoming quite noisy. In most cases, however, the fact that the data is
dropping below the early time curve is evident, pointing to the
necessity of a multi-exponential description. As before, these
predictions actually push quite deeply into the early time regime, but
the measurement window, and instrument dynamic range, are such as to
strongly limit the information content of the multi-exponential part of
the signal.

\section{Inverse problem for spheroidal targets}
\label{sec:inverse}


We have so far described applications of the early time and exponential
models to the \emph{forward problem,} in which detailed model
predictions for a known target are compared to data.  Having
demonstrated the quantitative success of the models for this problem,
we now turn to the \emph{inverse problem,} in which some set of target
characteristics is treated as unknown, and one attempts to determine
them by searching for the target model whose predictions best match the
data. We have previously presented solutions to the inverse problem
using noise corrupted simulated data \cite{WL03}. Here we will base the
inversions on the TEMTADS data.

Of particular interest are ambiguities in the data, i.e., target
properties that are poorly constrained by a particular data set, either
due to poor data quality (e.g., noise), or due to fundamental
trade-offs between certain parameters that exist even for essentially
perfect data. We will see, for example, that it is very difficult to
simultaneously determine target depth, size, and conductivity. We will
also see that the enormous differences seen between magnetic and
nonmagnetic target data in Sec.\ \ref{sec:datacomp} give rise to
similar differences for the inverse problem.

\subsection{Objective function}
\label{sec:objfn}

Inverse problems are generally formulated as the solution to the
following optimization problem. Let ${\bf m} \in M$ be a vector of
forward model parameters spanning a model space $M$, in our case the
space of target and sensor platform parameters that define the forward
problem. Let ${\bf d}$ be the measured data vector, in our case the set
of voltage time series corresponding to a given set
transmitter-receiver combinations and platform positions. Let ${\bf
d}_\mathrm{pred}({\bf m})$ be the forward model prediction for the data
given a model ${\bf m}$. For a perfect model and perfect data, there
should be a unique model ${\bf m}_0$ for which ${\bf
d}_\mathrm{pred}({\bf m}_0) = {\bf d}$. In the presence of noise,
and/or an incomplete model (e.g., real UXO are not ideal spheroids),
such an exact match is not achievable, and we instead seek an optimal
solution rather than a perfect solution.

The sense in which a solution is optimal is defined by an
\emph{objective function} $F({\bf m}|{\bf d})$, which, e.g., vanishes
if ${\bf d}_\mathrm{pred} = {\bf d}$, but more generally attains a
minimum value ${\bf m}_0$ in the space of available models:
\begin{equation}
{\bf m}_0 = \arg \min_{{\bf m} \in M} F({\bf m}|{\bf d}).
\label{5.1}
\end{equation}
Finding this minimum entails a search over the space $M$, and there may
be multiple local minima that interfere with the discovery of the
absolute minimum. It is generally the case, as well, that the problem
is under-determined, and an entire subspace of different ${\bf m}$, for
example, may very nearly achieve the same minimum. To some degree, such
ambiguities can be controlled using \emph{a priori} constraints (e.g.,
on a target's permitted range of depths or conductivities), effectively
biasing $F({\bf m}|{\bf d})$ toward a preferred set of solutions. The
danger, of course, is that this blinds one to targets disagreeing with
this bias, and ${\bf m}_0$ may end up being vastly different from the
truth. We will see examples of this below.

In what follows, we will use the following form of objective function:
\begin{equation}
F({\bf m}|{\bf d}) = \sum_{k=1}^{N_t}
W_k \left[\frac{V_\mathrm{meas}(t_k)}
{V_\mathrm{pred}(t_k;{\bf m})} - 1 \right]^2
+ F_\mathrm{prior}({\bf m}),
\label{5.2}
\end{equation}
in which $t_k$, $k=1,2,\ldots,N_t$ are the measurement platform time
gates, $V_\mathrm{meas}$, $V_\mathrm{pred}$ are the measured and
predicted voltages, and $F_\mathrm{prior}$ contains any prior
constraints. The weights $W_k$ may be used to fine tune the weighting
given to data in the different time regimes, for example suppressing
noise-dominated time gates. The ratio $V_\mathrm{meas}/V_\mathrm{pred}$
is used (adopting, basically, a logarithmic rather than a linear
voltage scale), in place, e.g., of a more conventional mean square
error $(V_\mathrm{meas} - V_\mathrm{pred})^2$, in order to give more
democratic weight to early and later-time regimes. Specifically,
signals may be orders of magnitude weaker at later time, but it will be
seen that the multi-exponential decay still contains key target
identification information not present at early time. In addition,
$V_\mathrm{pred}$ is placed in the denominator because noise effects
may easily induce sign changes in $V_\mathrm{meas}$ at later time,
which would produce singularities in $V_\mathrm{pred}/V_\mathrm{meas}$.

\begin{figure}

\centerline{\includegraphics[width=3.0in,bb=75 200 505 592,clip]
{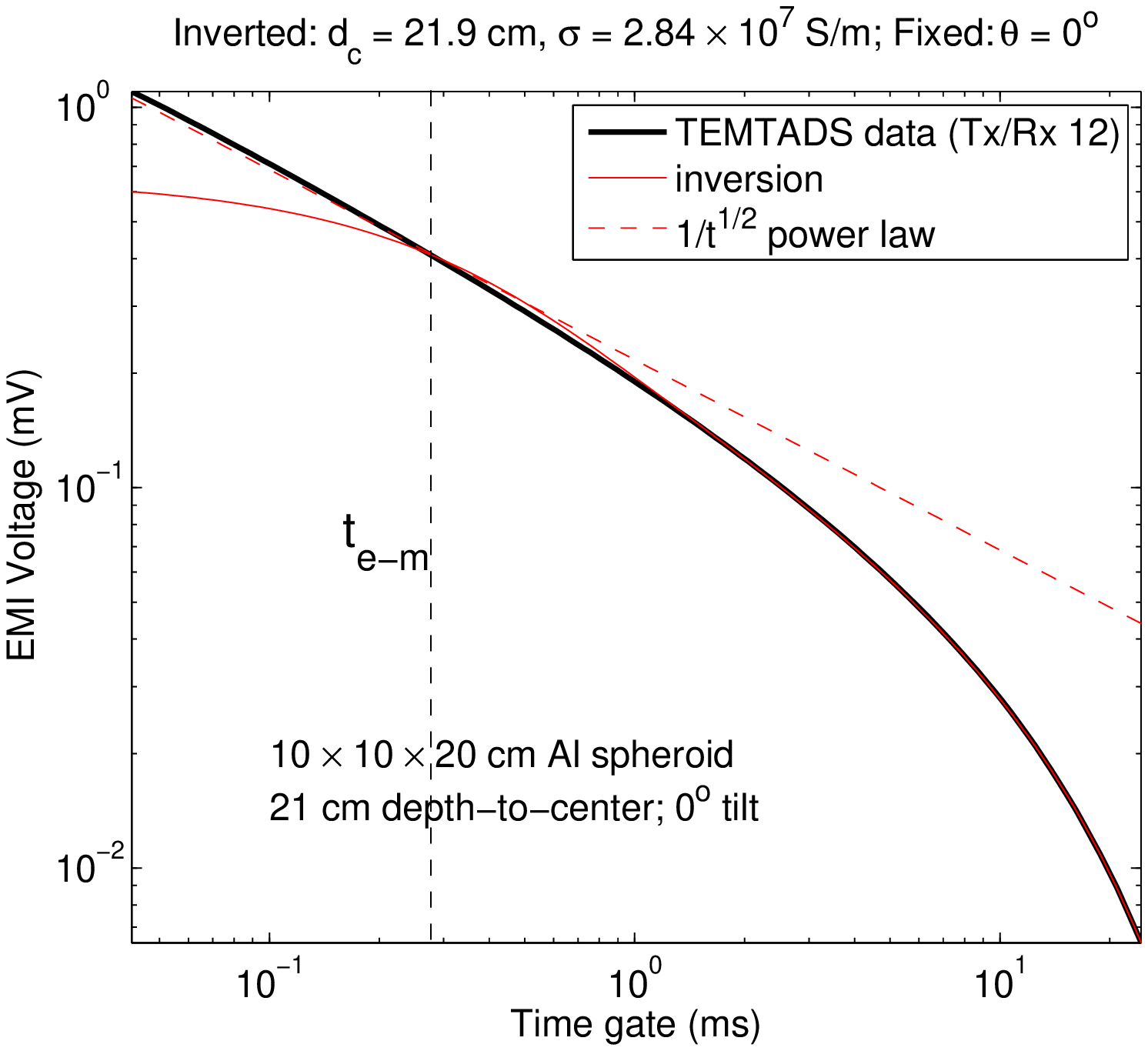}}

\smallskip

\centerline{\includegraphics[width=3.0in,bb=82 192 502 600,clip]
{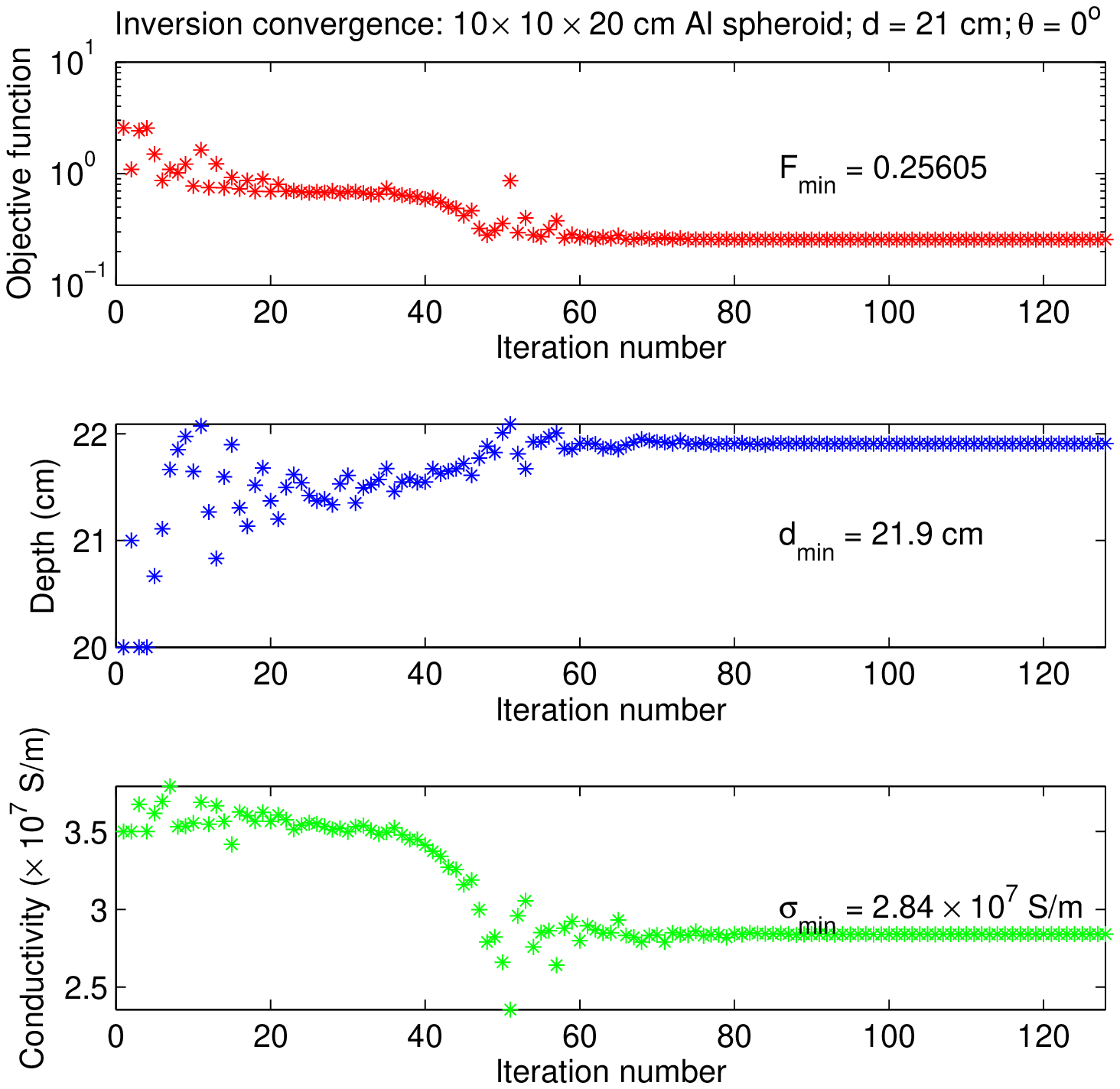}}

\caption{Inversion experiment using TEMTADS data for a $10 \times 10
\times 20$ cm prolate spheroid, using the objective function
(\ref{5.2}). Prior information is incorporated in that only the
parameters ${\bf m} = \{\sigma,d,t_{\mbox{e-m}} \}$ are permitted to
vary. Target geometry and orientation (symmetry axis vertical) are
fixed, and the transmitter current amplitude is fixed at 5.5 a. The
upper plot shows the data (thick black line---the same as the green
curve in Fig.\ \ref{fig:alprolate}) and the remarkably accurate optimal
fit (solid red line to the right of $t_{\mbox{e-m}}$; dashed red line
to its left). The lower plot shows the convergence of $F,\sigma,d$ with
iteration number. Note the two-stage convergence, in which $d$
equilibrates first, followed by the `weaker' parameter $\sigma$.}

\label{fig:Alinv1}
\end{figure}

\begin{figure}

\centerline{\includegraphics[width=3.0in,bb=75 195 507 602,clip]
{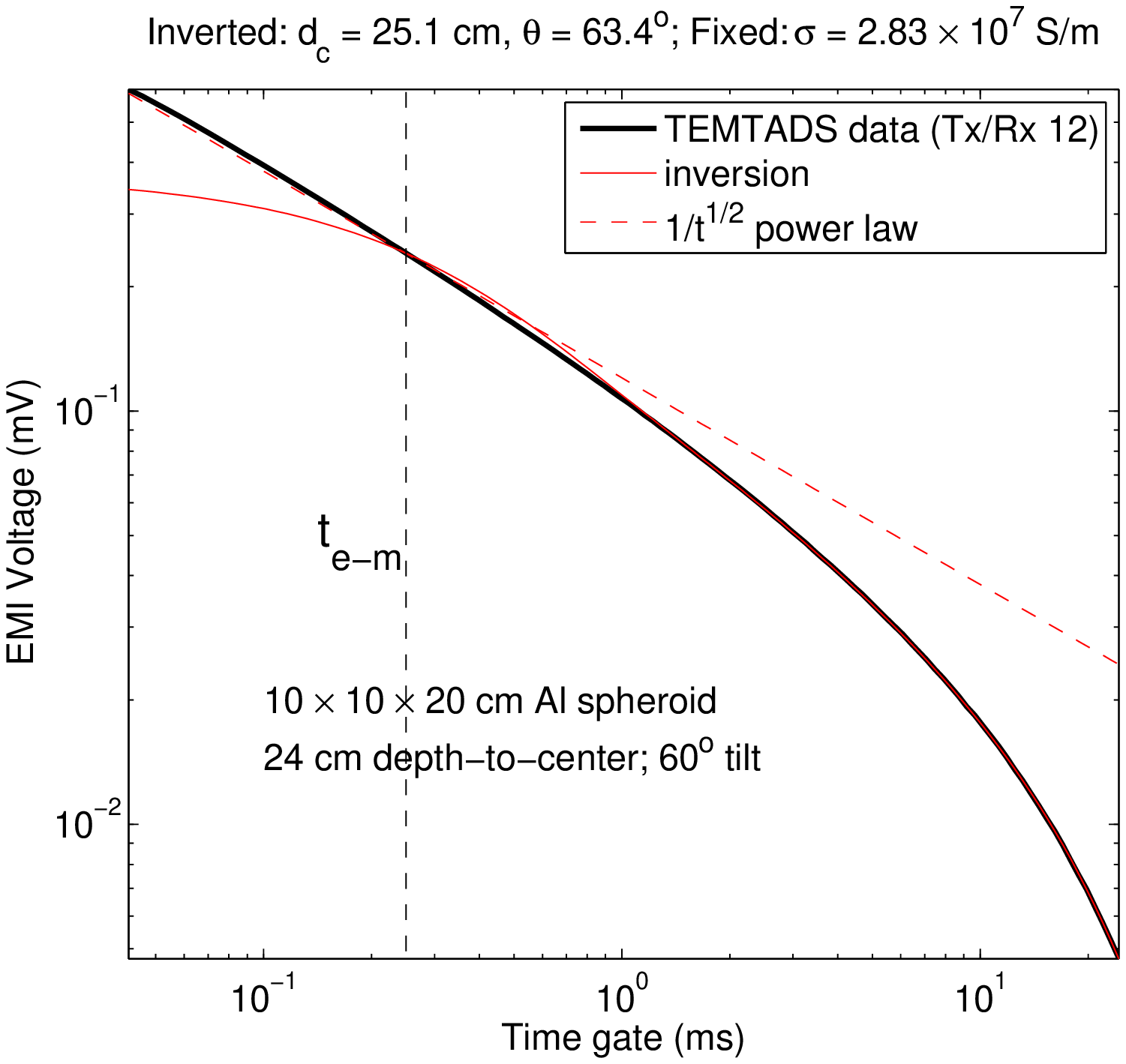}}

\smallskip

\centerline{\includegraphics[width=3.0in,bb=82 192 504 600,clip]
{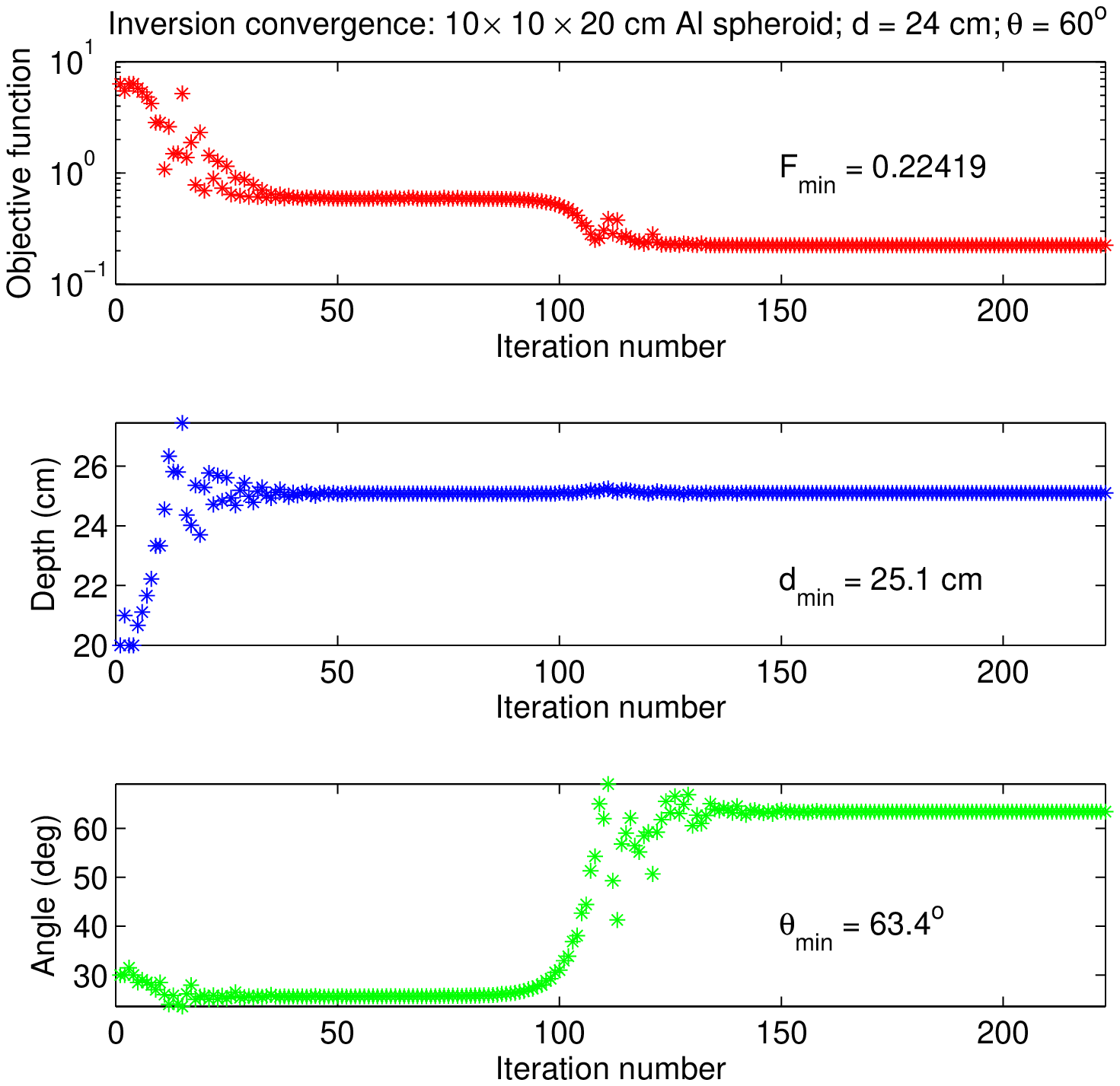}}

\caption{Second inversion experiment using TEMTADS data for a $10
\times 10 \times 20$ cm prolate spheroid, using the objective function
(\ref{5.2}). The parameters ${\bf m} = \{d,\theta,t_{\mbox{e-m}} \}$
are now permitted to vary. Target geometry is fixed, the transmitter
current amplitude is fixed at 5.5 a, and the conductivity is fixed at
the value found in the first experiment (Fig.\ \ref{fig:Alinv1}). The
upper plot shows the data (thick black line---the same as the magenta
curve in Fig.\ \ref{fig:alprolate}) and optimal fit (solid red line to
the right of $t_{\mbox{e-m}}$; dashed red line to its left). The lower
plot shows the convergence of $F,\theta,d$ with iteration number. The
value of $\theta_\mathrm{min} = 63.4^\circ$ is very close to the
$60^\circ$ ground truth. One sees even more clearly in this figure the
initial equilibration of the depth, followed later by the convergence
of the tilt angle, which has a much weaker effect on the voltage time
series, mostly at later time.}

\label{fig:Alinv2}
\end{figure}

\subsection{Inversions for aluminum targets}
\label{sec:Al_inv}

We begin with a set of numerical inversion experiments using the $10
\times 10 \times 20$ cm aluminum spheroid data (upper curves in Fig.\
\ref{fig:alprolate}). We consider some interesting issues involving
tradeoffs between target depth, conductivity, and orientation, which
are most clearly elucidated by treating the target geometry (in this
case, $a_{xy} = 5$ cm radius, aspect ratio $\alpha = a_z/a_{xy} = 2$)
as known. We have performed inversions in which the target geometry
also varies (see also Ref.\ \cite{WL03}) and found similar effects. We
will also use the (highest quality) data only from the pair Tx12-Rx12.
In Sec.\ \ref{sec:realUXO}, some results will be shown using multiple
Tx/Rx pairs. The inversion code uses the standard simplex method (which
`walks' its way toward a local minimum, in sequentially decreasing
steps), that does not require any gradients of the objective function.
More sophisticated search methods, that may operate more efficiently,
will be left for future work.

The transition between the early-time and multi-exponential regimes for
non-magnetic targets is treated as follows. One of the inversion
parameters is taken to be a crossover time $t_\mathrm{e\mbox{-}m}$, and
for a given value, the predicted voltage takes the form
\begin{equation}
V_\mathrm{pred}(t) = \left\{\begin{array}{ll}
V_\mathrm{MF}(t), & t > t_\mathrm{e\mbox{-}m} \\
V_\mathrm{MF}(t_\mathrm{e\mbox{-}m}) (t/t_\mathrm{e\mbox{-}m})^{-1/2},
& t < t_\mathrm{e\mbox{-}m},
\end{array} \right.
\label{5.3}
\end{equation}
in which $V_\mathrm{MF}(t)$ is the direct mean field prediction based
on the remaining parameters in ${\bf m}$ \cite{foot:etexact}.

In a number of numerical experiments, it was found that if the target
conductivity $\sigma$, the depth $d$, and the tilt angle $\theta$ are
all permitted to vary, the inversion is very unstable.  Specifically,
varying the tilt changes the signal decay rate at later time
(reflecting differences in the decay rates of the excited modes).
However, so does varying the conductivity, and the two effects can very
nearly be made to cancel, so long as the depth is adjusted slightly to
maintain the observed overall signal amplitude.  The result is a
solution ${\bf m}_0$ with unphysical values of all three parameters. In
realistic scenarios, one would then have to include in
$F_\mathrm{prior}$ terms that constrain, for example, the conductivity
to in a range of acceptable values for aluminum.

\begin{figure}

\centerline{\includegraphics[scale=0.6]{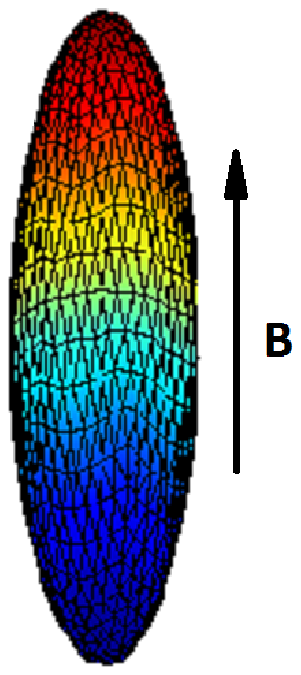} \quad
\includegraphics[scale=0.6,bb=0 -45 225 70]{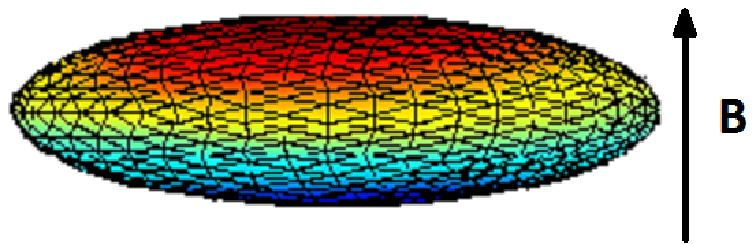}}

\caption{Early time surface current flows depending on relative
orientation of the magnetic field for a 4:1 prolate spheroid. Plotted
is the stream function $\psi$ associated with two of the early time
surface modes whose level curves are the stream lines of the surface
current (circulating clockwise around blue patches, counterclockwise
around red patches). The characteristic radius $r_\mathrm{eff}$
associated with a mode is essentially the (half) distance between
extrema of $\psi$ (red and blue regions). The diffusion of the surface
current sheet into the target, at any given point on the surface, is
shown in Fig.\ \ref{fig:earlyt_profiles} for different values of
$\kappa = (\mu/4\pi \sigma)^{1/2} c r_\mathrm{eff}$.}

\label{fig:earlytmodes}
\end{figure}

Since, for the data dealt with in this section, we have complete ground
truth, we proceed in a slightly different fashion. We begin by
\emph{determining} an optimal value of the conductivity from one of the
data sets using the known value of the tilt angle (as well as the
target horizontal position). The depth and crossover time are permitted
to vary as well. The results are shown in Fig.\ \ref{fig:Alinv1}, where
a value $\sigma = 2.84 \times 10^7$ S/m is found. Using a different
value of $\sigma$ is found to significantly change the inversion result
for the tilt angle.

Note that convergence of the inversion takes place in two stages. The
depth equilibrates first, due to the strong $1/d^6$ dipolar dependence
of the overall voltage amplitude. Once a reasonable value for $d$
emerges, the more subtle conductivity-induced changes in the voltage
curve shape (mainly at later time) can be productively optimized.

Note also that the inverted depth $d_\mathrm{min} = 21.9$ cm differs
slightly from the 21 cm ground truth value. This may partially be due
to experimental error, but is likely also due to the $\sim$10\%
uncertainty in the transmitter current amplitude $I_\mathrm{max}$ (see
Sec.\ \ref{sec:tx_wvfrm}) which has been fixed at the value 5.5 a.
Inversion experiments have been performed in which $I_\mathrm{max}$ is
also treated as a free parameter, but this is also found to be highly
unstable. Specifically, the voltage signal changes very little if
$I_\mathrm{max}$ is varied, while at the same time adjusting the target
depth $d$---both mainly affect the overall amplitude of the voltage
time series, not its shape \cite{foot:multiTxRx}. For the same reason,
here and in experiments described below, this ambiguity has very little
effect on the inverted values of most other parameters, specifically
those that indeed impact the shape of the time series.

With an accurate conductivity value now in hand, in our second
inversion experiment, we fix the former and invert for the tilt angle
\cite{foot:azimuth}. As seen in Fig.\ \ref{fig:Alinv2}, this indeed
produces an inverted $\theta_\mathrm{min}$ very close to the $60^\circ$
truth. The two-step nature of the convergence is seen here as well.
Similar experiments (not shown) using the $30^\circ$ and $90^\circ$
tilt data (red and cyan curves, respectively, in Fig.\
\ref{fig:alprolate}) produce equivalent results. The inverted depth
$d_\mathrm{min} = 25.2$ cm differs only slightly from the 24.4 cm
ground truth.

We have performed inversions as well using the $20 \times 20 \times 8$
cm oblate spheroid data shown in Fig.\ \ref{fig:aloblate}) which yield
very similar results (not shown). One again fits the conductivity (and
depth) using the $0^\circ$ tilt data (green curve in Fig.\
\ref{fig:aloblate}), and then uses this value to invert for tilt angle
(and depth) using the other data sets. One again finds delicate
features where a slightly incorrect conductivity value leads to a
highly inaccurate tilt angle. The problem actually appears to be
somewhat worse in this case because of the early time crossover time
$t_{e\mbox{-}m}$ occurs relatively earlier for oblate geometries,
increasing the model misfit for the given number of modes (232).

\begin{figure*}

\centerline{\includegraphics[width=3.2in]{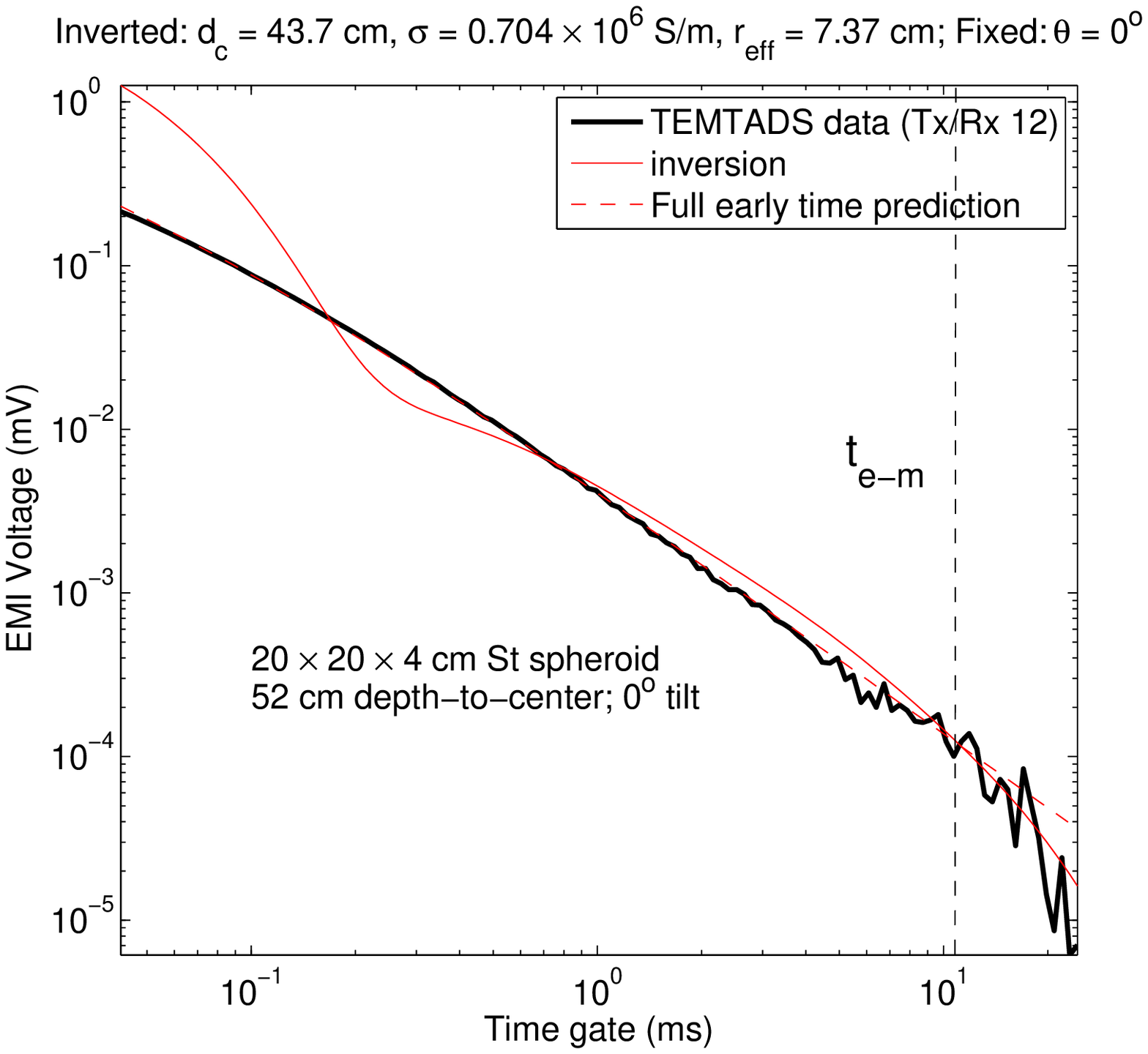} \quad
\includegraphics[width=3.2in]{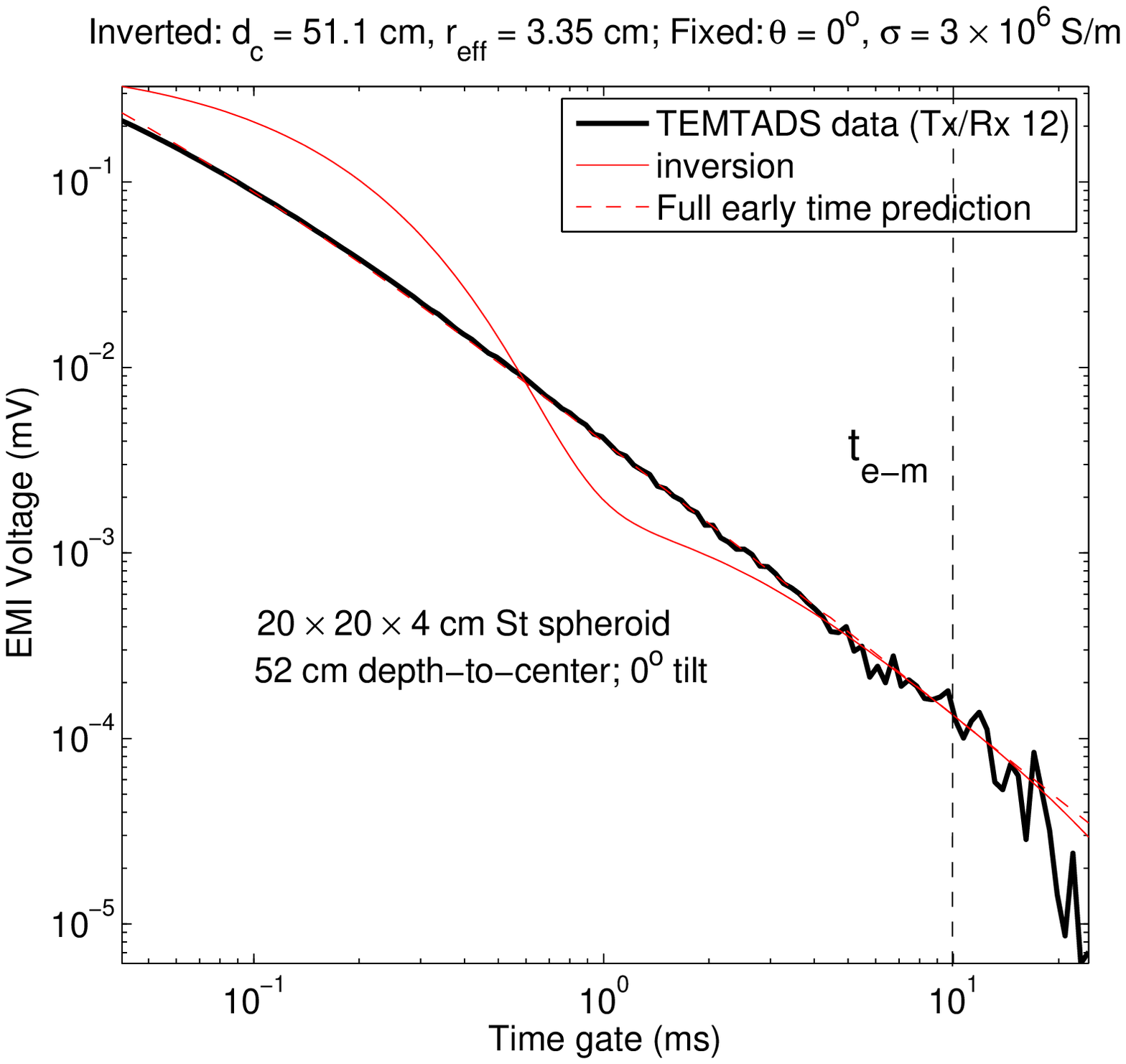}}

\centerline{\includegraphics[width=3.2in]{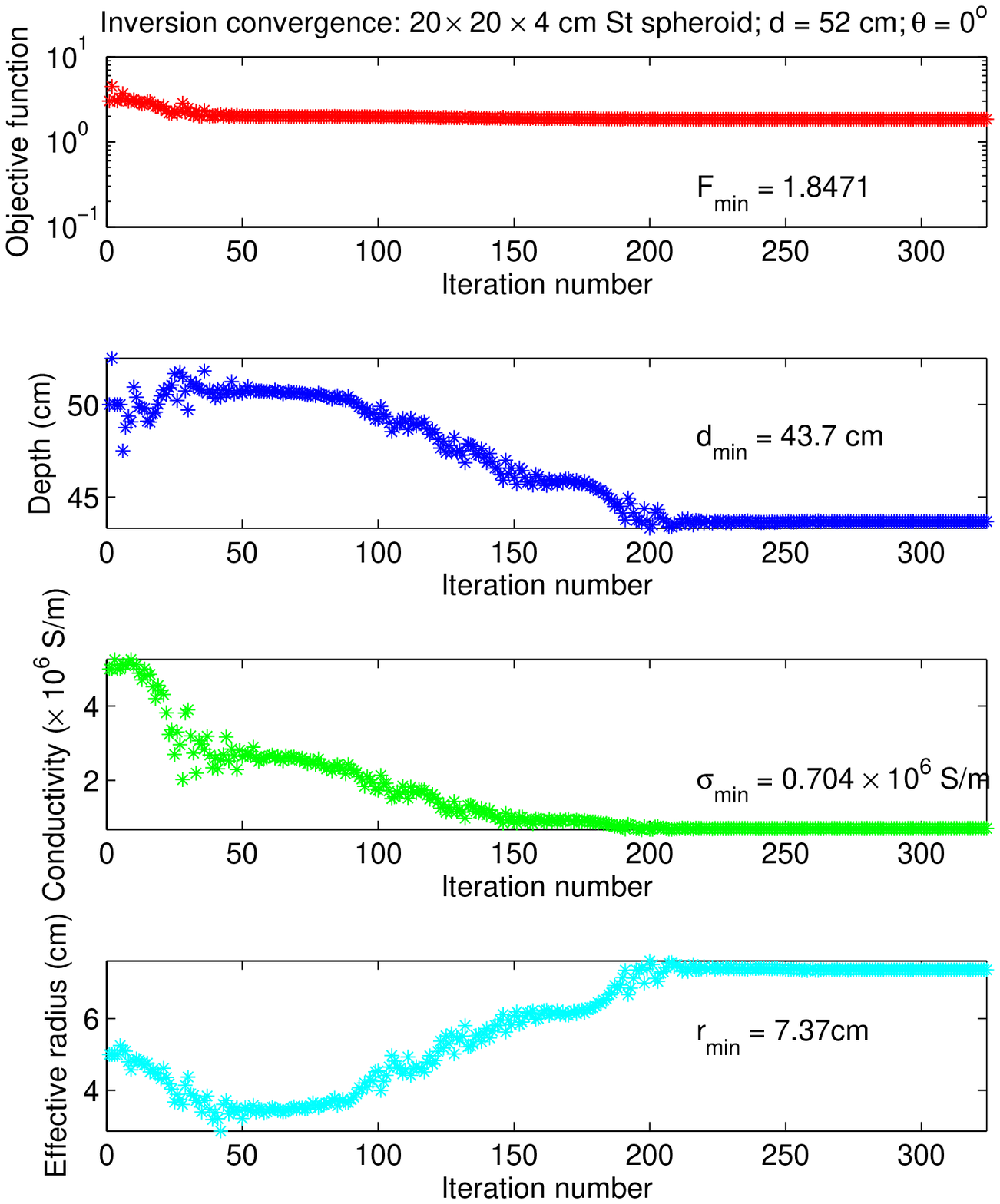}
\quad
\includegraphics[width=3.2in]{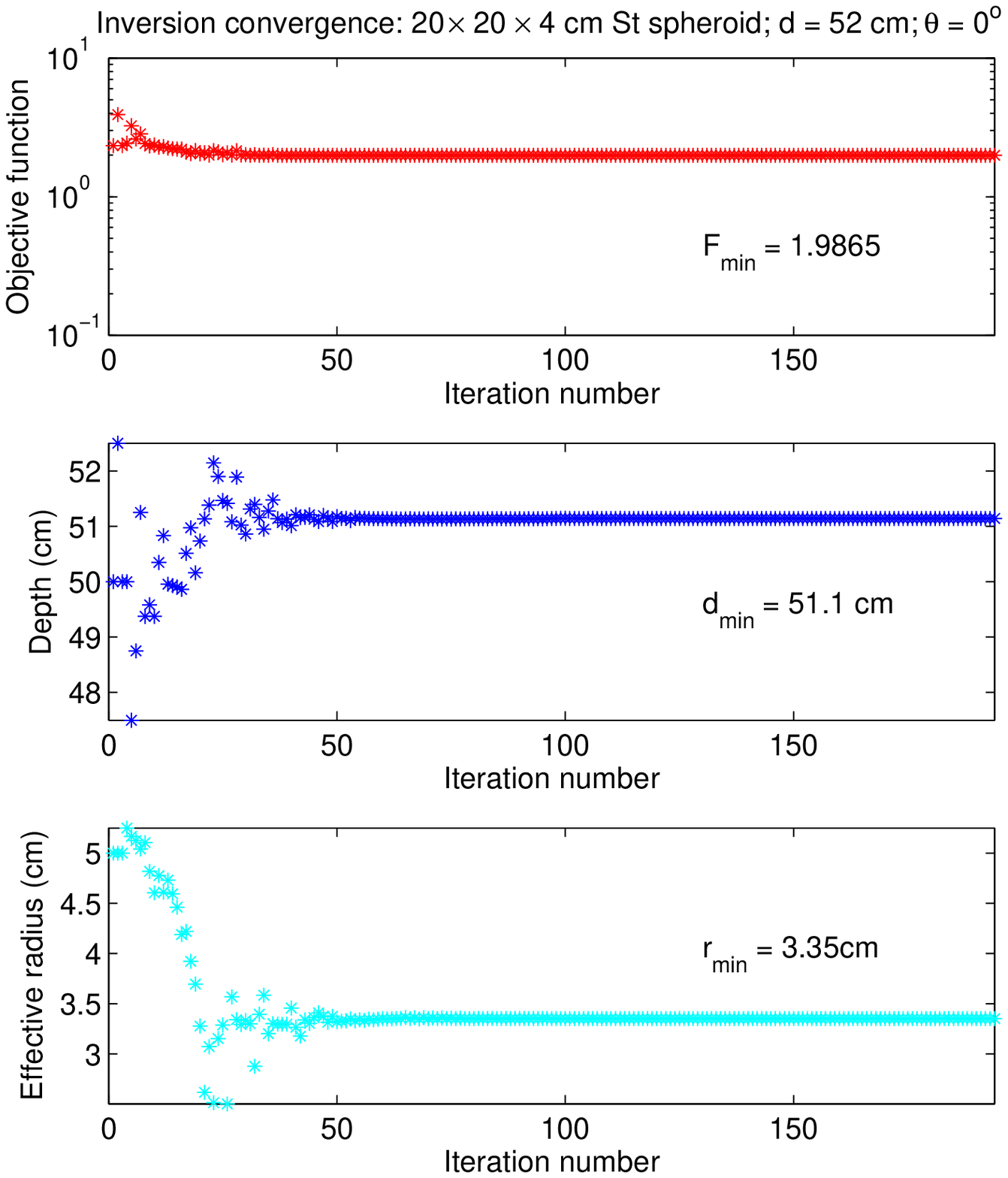}}

\caption{Two inversion experiments using TEMTADS data for a $20 \times
20 \times 4$ cm steel oblate spheroid, using the objective function
(\ref{5.2}). Rather than producing a simple power law, the early time
theory now produces the nontrivial `error function' series (\ref{2.11})
and (\ref{2.12}), modeled here as a single term with optimized value of
the effective radius $r_\mathrm{eff}$ (solid red line in the upper
plots). The target orientation is held fixed (symmetry axis vertical).
\textbf{Left hand plots:} the depth, conductivity, and effective radius
are permitted to vary. The resulting over-fit of the later-time noise
produces unphysical values of all three. \textbf{Right hand plots:} the
conductivity is now fixed at a physically reasonable value $\sigma = 3
\times 10^6$ S/m. The depth, conductivity, and effective radius now
converge to physically reasonable values as well.}

\label{fig:Stinv}
\end{figure*}

\begin{figure*}

\centerline{\includegraphics[width=3.2in,bb=45 165 540
625,clip]{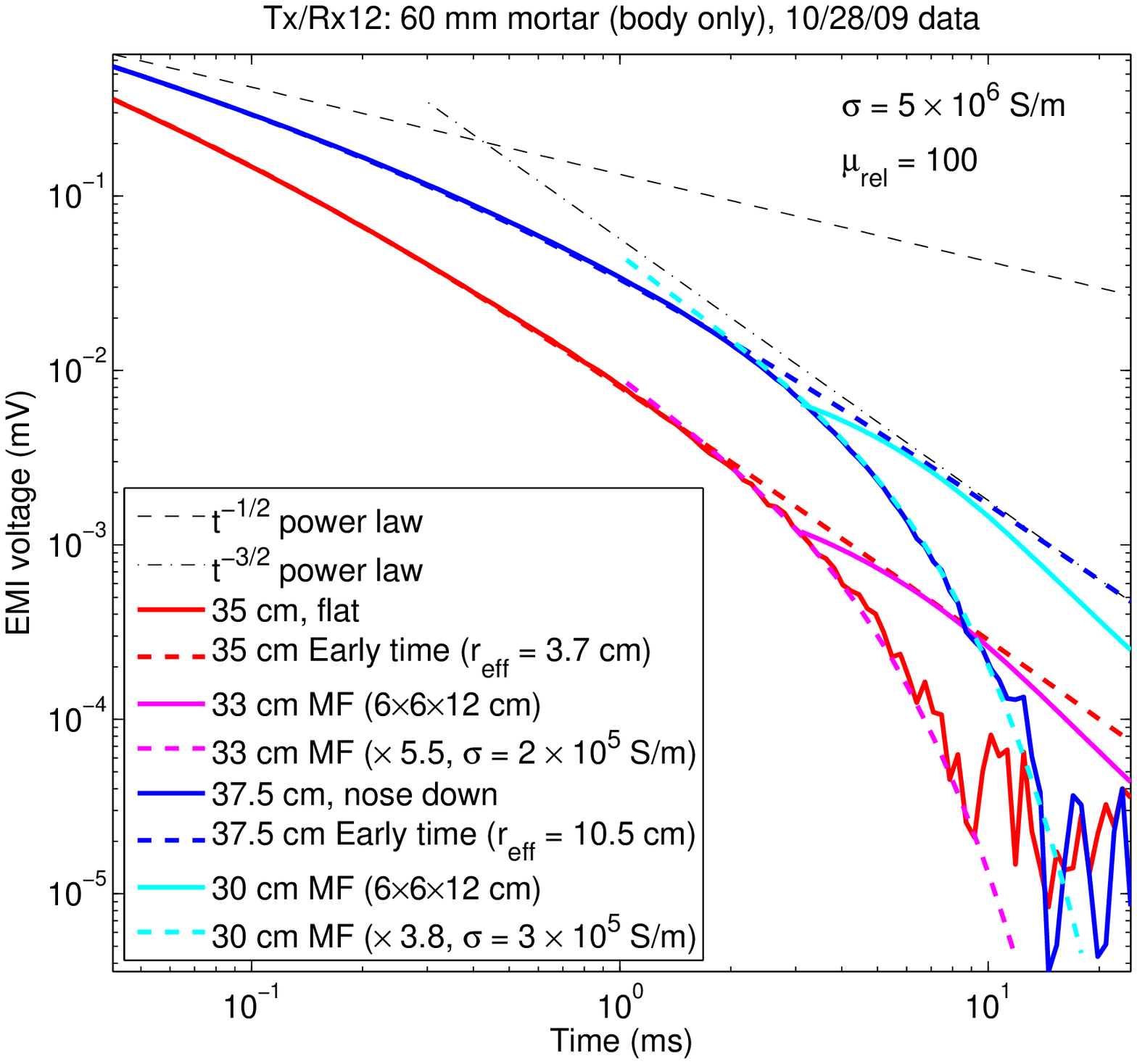} \quad
\includegraphics[width=3.2in,bb=45 165 540 625,clip]{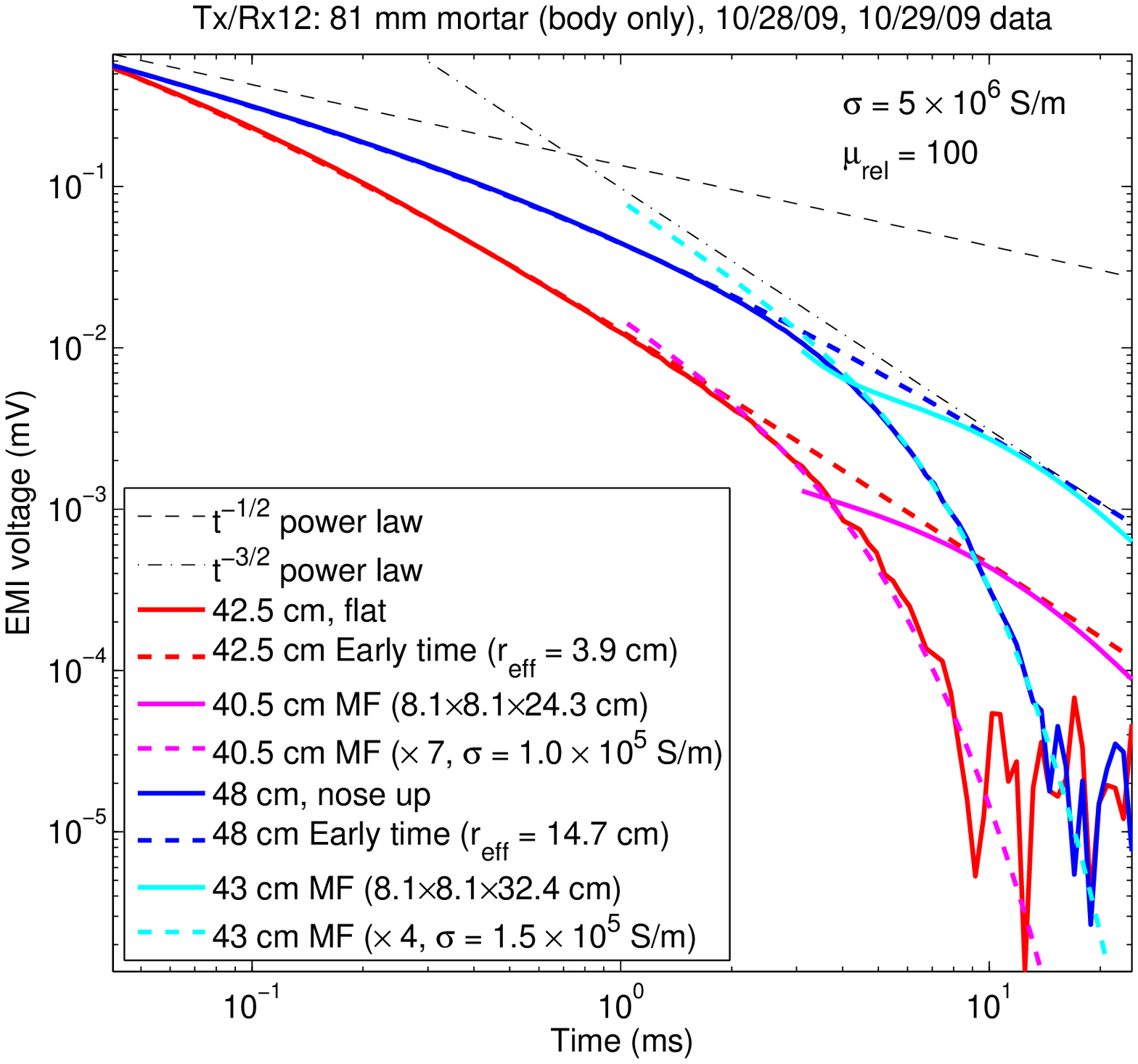}}

\caption{TEMTADS data (solid lines) and theoretical fits (dashed lines)
taken for a 60 mm (left) and 81 mm (right) mortar body targets lying
flat (red and magenta curves) and nose-up vertical (blue and cyan
curves). The early time fits, based on (\ref{5.4}) and (\ref{5.5}) with
$\mu = 100$ and $\sigma = 5 \times 10^6$ S/m, produce the very
reasonable effective diameters $d_\mathrm{eff} = 2r_\mathrm{eff}$ (see
the illustrations in Fig.\ \ref{fig:earlytmodes}) as indicated in the
legends. The length listed at the beginning of each legend entry is the
target depth (ground truth for the data and early time curves; fitted
value for the mean field curves). The mean field predictions (solid
magenta and cyan curves), using these same parameters, properly extend
the early time curves but completely fail to fit the data. Lowering the
conductivity by more than an order of magnitude accurately mimics the
increased rate of decay (dashed magenta and cyan curves), but the
overall voltage amplitude is incorrect (as indicated by the applied
multipliers in the legend). As described in the text, and in Fig.\
\ref{fig:hollowsph}, the correct explanation for the discrepancy is the
finite (0.5--1 cm) mortar shell thickness.}

\label{fig:mortars}
\end{figure*}

\subsection{Inversions for steel targets}
\label{sec:Stinv}

We next present inversions based on the data shown in Fig.\
\ref{fig:stoblate} for steel oblate spheroidal targets. These turn out
to serve as an illustration of another set of inversion pitfalls one
may encounter, in this case depending on how one balances prior
knowledge in the presence of noisy data in the later time regime.

Since the early time behavior of ferrous targets is much more
complicated than that for nonmagnetic targets, in addition to the
crossover time $t_\mathrm{e\mbox{-}m}$, one must provide an appropriate
parametrization of the magnetic surface mode series (\ref{2.11}) and
(\ref{2.12}) \cite{W2004}. As described in Sec.\ \ref{sec:earlyt},
replacing the series by a single term with effective parameters is
found to provide the best fit. For reasons that will become evident
below, we parameterize it in the form
\begin{equation}
V_e(t) = V_0^e H(\sqrt{t/\tau_e}),
\label{5.4}
\end{equation}
where
\begin{equation}
\tau_e = \frac{1}{\kappa^2}
= \frac{4\pi\sigma}{\mu c^2} r_\mathrm{eff}^2,
\label{5.5}
\end{equation}
with $r_\mathrm{eff}$ an effective early time target radius. This
parameter, derived as previously from the early time eigenvalue
$\kappa_n$, reflects the primary length scale of the surface currents
for a given mode, and decreases as $n$ increases and the geometric
complexity of the mode increases. This length scale may be thought of
as the diameter of the target along the direction of the magnetic field
that primarily serves to excite the mode. For example, currents
circulating around the symmetry axis $a_z$ are generated by a vertical
magnetic field (left panel of Fig.\ \ref{fig:earlytmodes}), and
$r_\mathrm{eff}$ is then found to be comparable to $a_z$. For magnetic
field orthogonal to $a_z$ the currents circulate up one side of the
symmetry axis and down the other (right panel of Fig.\
\ref{fig:earlytmodes}), and $r_\mathrm{eff}$ is found to be comparable
to the radius $a_{xy}$.

As a consequence the length $a_z$ will be reflected in the early time
data when $a_{xy}$ is what would observed in a corresponding visual
inspection of the target from the point of view of a
transmitter-receiver above. Correspondingly, when the target is laid
with symmetry axis horizontal, $a_{xy}$ will be exhibited when a visual
inspection would clearly see both $a_{xy}$ and $a_z$. These
observations will be confirmed by the data below.

Analogous to the inversion shown in Fig.\ \ref{fig:Alinv1}, in Fig.\
\ref{fig:Stinv} we show inversions for the parameters $d,\sigma,
t_\mathrm{e\mbox{-}m}, r_\mathrm{eff}$ with tilt angle fixed at the
$0^\circ$ ground truth value, and $\mu = 100$. As can be seen, the
crossover time $t_\mathrm{e\mbox{-}m} \simeq 10$ ms lies in a regime
where noise effects are becoming significant. In particular, in this
case the noise happens to induce a sharper downturn in the signal in
the multi-exponential regime than is case for the true signal. In
attempt to fit this, the inversion on the left unphysically suppresses
the conductivity ($\sigma_\mathrm{min} = 7.04 \times 10^5$ S/m), and
compensates with an unphysically large effective radius
($r_\mathrm{min} = 7.37$ cm, much larger than the 2 cm half-thickness
of the discus). This keeps the product $\sigma r_\mathrm{eff}^2$ in
(\ref{5.5}) essentially fixed, thereby maintaining a good fit in the
early time regime. Note that the depth $d_\mathrm{min} = 43.7$ cm also
differs substantially from the 52 cm ground truth.

In order to avoid this overfit of the noise, the right hand side of
Fig.\ \ref{fig:Stinv} shows the superior result obtained by fixing the
conductivity at the physically reasonable value $\sigma = 3 \times
10^6$ S/m, with the results $r_\mathrm{min} = 3.35$ cm and
$d_\mathrm{min} = 51.1$ cm.

The main lesson here is that significantly different information
(specifically, different combinations of conductivity with other
parameters) is contained in the early time and later time regimes for
ferrous targets, and there are large parameter ambiguities in the
absence of good data in both. For ferrous targets, the more rapid
$1/t^{3/2}$ dominant decay in early time leads to much smaller signals,
hence degraded data, in the multi-exponential regime. Parameters
relying on the latter will therefore be more poorly determined, and one
may be forced to apply a larger set of prior knowledge constraints than
for nonmagnetic targets. Since real UXO targets are usually ferrous,
this lesson has important practical implications.

\section{Some results for Real UXO targets}
\label{sec:realUXO}

We turn finally to some initial inversion results for real UXO targets,
namely 60 mm and 81 mm mortar bodies, selected for their
near-spheroidal shape. The former has a 6 cm diameter, and is
approximately 13 cm long; the aspect ratio is therefore taken as
$\alpha = 2$. The latter has an 8.1 cm diameter and is approximately 25
cm long; the aspect ratio is therefore taken as $\alpha = 3$. Real UXO
are not ideal spheroids, but we will see that key target discrimination
information can be obtained by comparing their electromagnetic
responses to those of spheroids with similar geometry.

Data and theoretical fits for both UXO are shown in Fig.\
\ref{fig:mortars}. The most important observation is that, with the
assumptions $\sigma = 5 \times 10^6$ S/m and $\mu = 100$, the early
time curves for the two orientations (vertical and flat) produce
excellent fits, and estimates for the effective radius that accord very
nicely with the discussion in Sec.\ \ref{sec:Stinv} (and Fig.\
\ref{fig:earlytmodes}): for the vertically oriented mortars,
$r_\mathrm{eff} = 10.5,\ 14.7$ cm are indeed comparable to the mortar
half-lengths, while for `flat' mortars, $r_\mathrm{eff} = 3.7,\ 3.9$ cm
are comparable to their radii. Note that these inferences are
independent of the target depth (which, to leading order, affects only
the amplitude $V_0^e$).

The next observation is that, contrary to the early time fit, the mean
field predictions (dashed blue and red lines in Fig.\
\ref{fig:mortars}) fail to fit the data. Thus, although they provide a
logical extension of the early time curves \cite{foot:targetdepths},
the data follow a much more steeply decaying path.  As indicated by the
red and magenta dashed lines, the decay can be fit using much smaller
(by factors of 20--50) conductivities. However, these unphysically
small effective values are also unphysically orientation dependent, and
give similarly inconsistent values for the overall voltage scales
(hence the $\times 4$--$\times 7$ fudge factor multipliers listed in
the figure legends).

\begin{figure}

\centerline{\includegraphics[width=3.2in,bb=80 205 500 580,clip]
{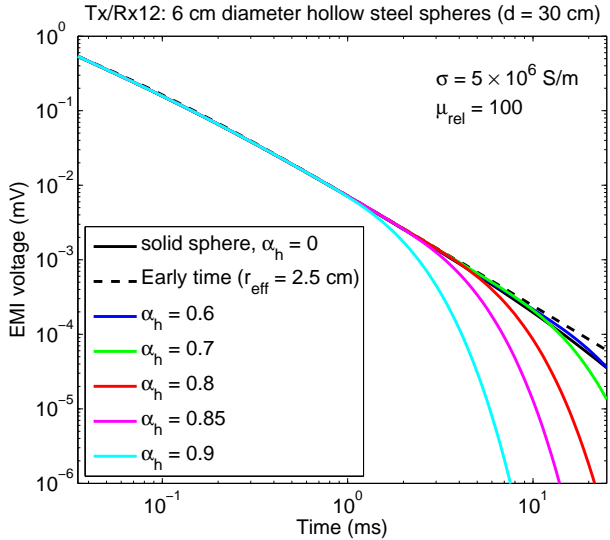}}

\caption{Exact analytic TDEM voltage predictions for a sequence of
hollow steel spheres, including the solid sphere ($\alpha_h=0$, solid
black line). Since the surface geometry of all targets is identical,
the same early time curve (dashed line) fits all data sets. However,
the multi-exponential regime begins earlier for thinner-shelled
targets.}

\label{fig:hollowsph}
\end{figure}

The origin of these modeling discrepancies is that the finite mortar
shell thickness (in the 0.5--1 cm range) has not been accounted for. To
illustrate this, in Fig.\ \ref{fig:hollowsph} exact analytic results
for a hollow sphere are shown. The results there confirm the rapid
increase in the mode decay rates with decreasing shell thickness (left
panel), and the resulting early breakaway of the voltage curves from
the early time result (right panel). The breakaway time scales with the
shell thickness as $t_X \sim w_\mathrm{shell}^2/D$, where $D = c^2/4\pi
\mu \sigma$ is the electromagnetic diffusion constant (see Sec.
\ref{sec:earlyt}). With the given values, this relation may be put in
the form:
\begin{equation}
\frac{t_X}{1\ \mathrm{ms}} \sim
\left(\frac{w_\mathrm{shell}}{1.3\ \mathrm{mm}} \right)^2.
\label{6.1}
\end{equation}
In Fig.\ \ref{fig:mortars}, ones observes $t_X \sim 2$--3 ms, which
indeed places $w_\mathrm{shell}$ in the right range (and is consistent
as well with the $\alpha_h = 0.8$, 0.85 curves---hence
$w_\mathrm{shell} = 0.3$--0.5 cm---in Fig.\ \ref{fig:hollowsph}).

To summarize, the early time fits provide direct estimates of the
target geometry, while the breakaway time $t_X$ away from the early
time extrapolation (as well as the solid spheroid mean field
prediction), provide direct estimates of the UXO shell thickness. These
are key parameters in target identification. Quantifying the latter
would require an enhancement of the mean field code to deal explicitly
with hollow spheroids. This can be done with modest effort, and will be
considered for future work.

\begin{figure*}

\centerline{\includegraphics[width=3.2in,bb=40 190 545 605,clip]
{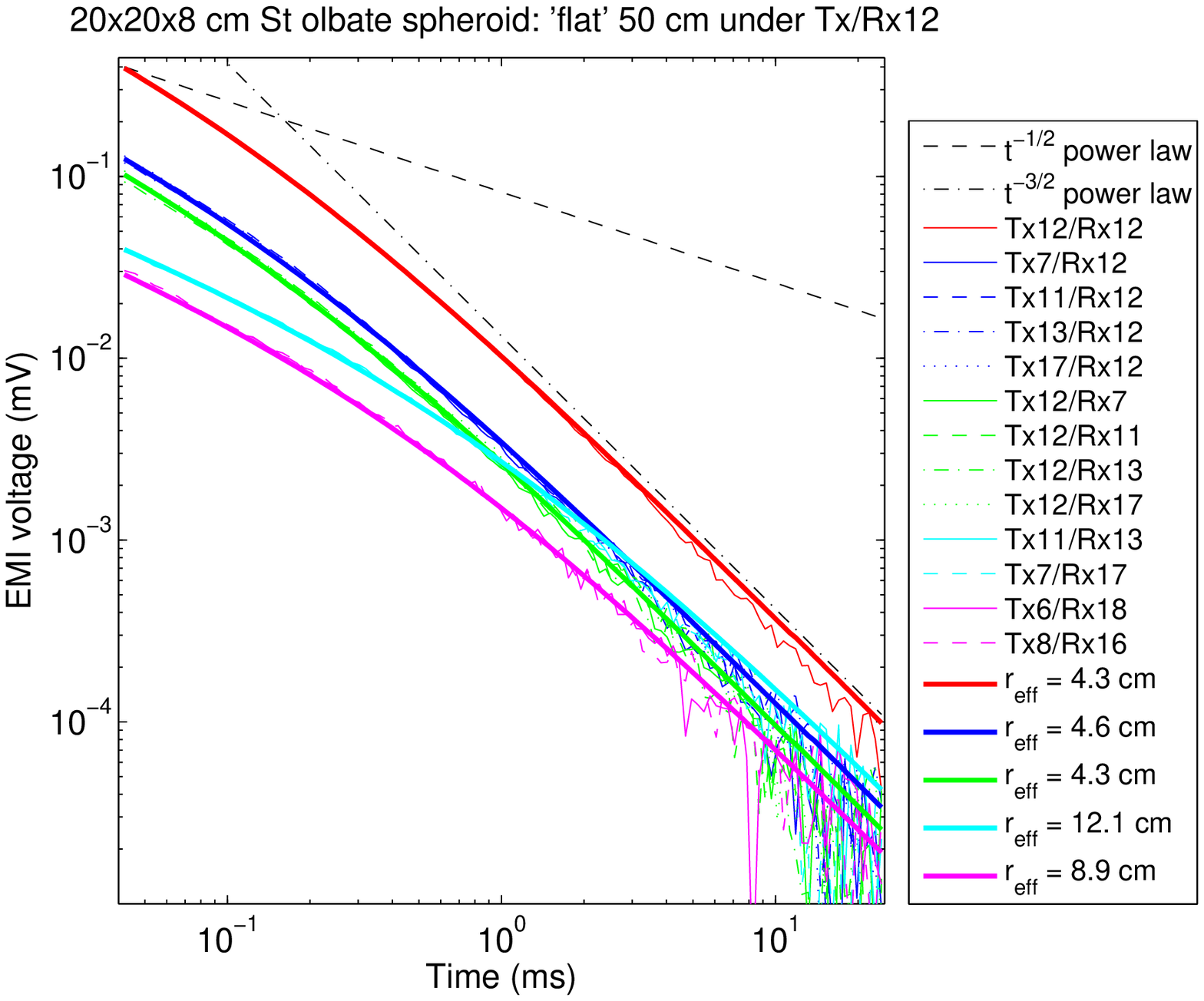} \quad
\includegraphics[width=3.2in,bb=40 190 545 605,clip]
{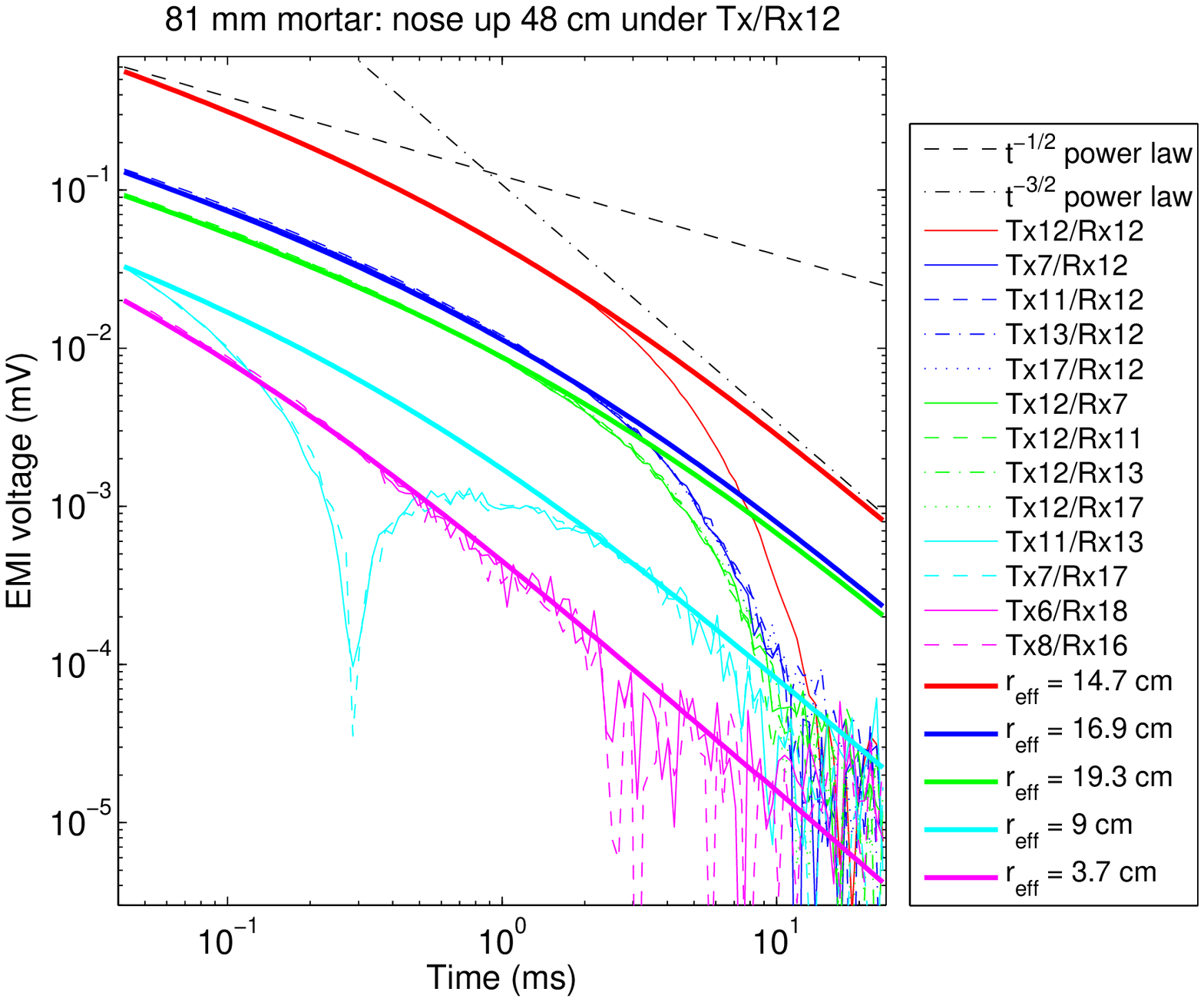}}

\caption{Data (thin solid and dashed lines) and early-time model fits
(thick solid lines) for two fixed steel targets using different Tx/Rx
combinations (with labeling shown in Fig.\ \ref{fig:temtads_array}).
\textbf{Left:} $20 \times 20 \times 8$ cm oblate spheroid.
\textbf{Right:} 81 mm mortar body. For these symmetrically placed
targets, the solid and dashed curve pairs of each color (e.g., Tx7/Rx12
and Tx11/Rx12; Tx11/Rx13 and Tx7/Rx17) should be degenerate, and this
is indeed observed. In both cases the early-time effective target
radius (listed in the legend) is seen to vary with `look angle' in a
way consistent with the target and measurement geometry. Note that the
breakaway time (\ref{6.1}) continues to be observable in a number of
the 81 mm voltage traces (and continues to be notably absent in the
solid spheroid traces). The cusp-like feature in the 81 mm Tx11/Rx13
and Tx7/Rx17 traces (solid and dashed cyan curves) actually represents
a sign change in the voltage, as discussed in the text.}

\label{fig:multiRxTx}
\end{figure*}

\subsection{Information content of multiple Tx-Rx combinations}
\label{sec:multit}

As described in the previous subsection, the early time prediction
provides an effective orientation-dependent target radius. For a given
buried target, the orientation is fixed, and one seeks other ways of
extracting this information. Here one may take advantage of the array
degrees of freedom available with the TEMTADS platform: each Tx-Rx pair
effectively provides a different `look' at the target. In Fig.\
\ref{fig:multiRxTx} we show such data for $20 \times 20 \times 8$ cm
oblate steel spheroid (left panel), and the 81 mm mortar (right panel).
In each case the targets are centered (50 cm and 48 cm, respectively)
below Rx/Tx12, with their symmetry axes vertical. See Fig.\
\ref{fig:temtads_array} for the labeling.

Note that if the transmitter and receiver coils were identical then, by
complementarity, the Tx(m)/Rx(n) response would coincide with that of
Tx(n)/Rx(m). The fact that they are somewhat different (see Table
\ref{table:temtads_array}) explains, for example, the small differences
between the sets of green and blue curves (e.g., Tx7/Rx12 vs.\ Tx12/Rx7
or Tx17/Rx12 vs.\ Tx12/Rx17). On the other hand, if two sets of Rx/Tx
pairs are symmetrically placed relative to the (axially symmetric)
target, then the responses should be identical (e.g., Tx7/Rx12 vs.\
Tx11/Rx12 or Tx11/Rx13 vs.\ Tx7/Rx17), and this is indeed observed.

To leading order, the actual time series is a superposition of the
horizontal and vertical early time magnetic surface modes (Fig.\
\ref{fig:earlytmodes}), and one expects that a fit to a single value of
$r_\mathrm{eff}$ will then find values that interpolate between the
two. This is indeed, for the most part, seen to be the case. For the
combination Rx12/Tx12 (red lines), the effective radius is indeed
comparable to the `vertical half-height' of the target (4 cm, vs.\
$r_\mathrm{eff} = 4.3$ cm for the oblate spheroid; 12 cm vs.\
$r_\mathrm{eff} = 14.7$ cm for the 81 mm mortar).

So long as either Rx12 or Tx12 is used (e.g., green and blue curves in
Fig.\ \ref{fig:multiRxTx}), $r_\mathrm{eff}$ remains comparable to (or
even larger than) its Rx12/Tx12 value. For Tx12, this is because the
same dominant magnetic surface mode is being excited, irrespective of
which receiver is used to observe it. For Rx12 the same argument
proceeds through complementarity: an off-center transmitter will excite
more than one surface mode, but the symmetrically placed receiver will
see only the symmetrically circulating mode.

However, when both transmitter and receiver are off-center (e.g., cyan
and magenta curves in the figure), the effective radius is seen to
become comparable to the physical (horizontal) radius of the target (10
cm for the oblate spheroid; 4.05 cm for the mortar). Thus, despite the
fact that the overall signal levels drop precipitously as the Tx-Rx
separation increases, to the point where the multi-exponential regime
becomes unobservable \cite{foot:txobs}, robust target geometry
inferences can still be made using this data through the early-time
prediction fits. It should also be noted that these fits, being
restricted to the early time regime, entail adjustment of the single
parameter $r_\mathrm{eff}$ (or $\kappa$), and so do not require a
sophisticated inversion scheme.

There is one feature of the 81 mm mortar data that deserves further
comment. The Tx11/Rx13 and Tx7/Rx17 (cyan curves in the left panel of
Fig.\ \ref{fig:multiRxTx}) display an unusual cusp feature at about 0.3
ms. Since the magnitude of the voltage is being plotted, this actually
corresponds to a node in the response, i.e., a sign reversal of $V(t)$.
The easiest way to understand this effect is to note that a target
close to Tx12 will generate a net downward-pointing magnetic field
through Rx13. However, as the target depth increases, the field becomes
net upward-pointing, reversing the sign of the flux, and hence of the
induced voltage. In this picture, a node in the response occurs as a
function of target depth, but a similar argument would also produce a
node as a function of time due to exchange of dominance of two
early-time mode contributions. Vertical variation of the applied
magnetic field will always produce higher order modes (with more
complex spatial structure than those shown in Fig.\
\ref{fig:earlytmodes}) with faster decay rates, and near the critical
depth the multiple contributions sum to produce a zero crossing.

\section{Summary and conclusions}
\label{sec:conclusions}

The results presented in this paper demonstrate the unprecedented
accuracy available from our first principles, physics based models
covering the entire measurement window, from the early time multi-power
law regime, all the way through the multi-exponential regime to the
late time mono-exponential regime. Prior to the mean field code's
current iteration \cite{W2011,W2012}, the number of accurately computed
modes used to describe the multi-exponential regime was limited to
perhaps a few dozen \cite{WL03}. As seen in the results presented, by
generating the required overlap of the early time and multi-exponential
regimes, this improvement is critical to the success of the validation
and inversion tests.

It should be emphasized that the increase in predictive power continues
to operate with extremely high numerical efficiency. The creation of
the mode data for a given target cannot be performed in real time, but
once this data is made available in a database that spans the expected
target geometries, its acquisition and use for measurement predictions
can be performed in real time---operating at essentially the same speed
as predictions using the exact solution for the sphere.

As seen in the figures, the dominant regimes visible in the data depend
very strongly on the target size and physical properties. Increasing
target size and magnetic permeability expands the early time regime to
later physical time. Smaller aluminum targets (e.g., blue lines in
Fig.\ \ref{fig:alprolate}) are completely described by the mean field
approach over the full time range, while even the smaller steel targets
barely enter multi-exponential regime (see Figs.\ \ref{fig:stprolate}
and \ref{fig:stoblate}) before the signal fades into the noise floor
\cite{foot:noise}.

Through investigation of inversions it has been shown that different
time regimes and different Tx/Rx pairs contain complementary target
discrimination information. The most exciting development is the
observation that the early time regime, which dominates the steel
target response, contains direct information about the target geometry
(via the different effective target diameters seen from different `look
angles') and the hollow target shell thickness (through the earlier
breakaway time to multi-exponential decay for thinner shelled targets).
These are key features that will support target identification and
clutter rejection.

All of these features, whose quantitative interpretation is enabled by
the present models, can be applied to the pursuit of robust target
discrimination and identification under more challenging conditions.
The code efficiency becomes especially critical for this purpose, as
searches through the database for the target whose response best
matches the data requires hundreds, or perhaps even thousands, of
iterations. In the inversion results presented, these searches
presently take several minutes. Further focused improvements in code
efficiency could probably reduce this to under one minute.

\section*{Acknowledgment}

This material is based upon work supported by SERDP, through the US
Army Corps of Engineers, Humphreys Engineer Center Support Activity
under Contract No.\ W912HQ-09-C-0024. The author has greatly benefitted
from discussions with D. Steinhurst, E. H. Hill, and E. M. Lavely.

\ifCLASSOPTIONcaptionsoff
  \newpage
\fi




\begin{thebibliography}{1}
\bibitem{foot:multiscale} This is far less obvious on the linear-time
    plot, in which only the $1/t^{3/2}$ late-early time power law is
    clearly visible. The log-time scale is key to elucidating the
    multi-scale nature of the target electrodynamics.
\bibitem{foot:multiTxRx} Simultaneous inversions of multiple time
    series would probably remove most of this particular ambiguity,
    since different Tx-Rx pairs provide different `looks' at a target
    that vary significantly relative to one another with depth.  This
    will be a subject of future work.
\bibitem{foot:azimuth} Target orientation, of course, is determined by
    the azimuthal angle $\phi$ as well. However, for a target centered
    under the Tx/Rx coil, the dependence on $\phi$ is extremely weak
    (and would be nonexistent for circular coils), and is therefore
    treated as fixed at the experimental value (along the line joing
    the centers of Rx/Tx10 and Rx/Tx/14---see Fig.\
    \ref{fig:temtads_array}). Simultaneous inversions of multiple Tx-Rx
    pairs would effectively remove this ambiguity as well (cf.\
    footnote \cite{foot:multiTxRx}).
\bibitem{foot:targetdepths} Note that for the targets lying flat, the
    estimated target depths lie only 2 cm from their ground truth
    values, while for vertical targets there are much larger (7.5 cm
    and 5 cm) discrepancies. One possible reason for this is the
    signicant nose-tail asymmetry of the mortars, which will generally
    bias the effective spheroid center away from the target geometric
    center.
\bibitem{foot:txobs} Though, for the 81 mm mortar this regime begins
    earlier, and the accompanying breakaway time $t_X$, equation
    (\ref{6.1}), continues to be apparent in a number of the voltage
    traces in the right panel of Fig.\ \ref{fig:multiRxTx}.
\bibitem{foot:noise} This signal fading is exacerbated by the more
    rapid $1/t^{3/2}$ decay of the late-early time signal for steel
    targets, as compared to the much slower $1/t^{1/2}$ aluminum
    targets. The latter then have greater tendency to maintain strong
    signals into the multi-exponential regime (see Figs.\
    \ref{fig:alprolate} and \ref{fig:aloblate}). As decribed in Sec.\
    \ref{sec:earlyt} (see especially Fig.\ \ref{fig:earlyt_profiles}),
    this more rapid decay has its orgin in the surface current
    dynamics, which, in magnetic targets, tends to more quickly push
    the currents away from the target surface.
\end{thebibliography}
%

%


\begin{IEEEbiographynophoto}{Peter B. Weichman}
is a principal research scientist at BAE Systems, Advanced Information
Technologies.
\end{IEEEbiographynophoto}






\end{document}